\documentclass[aps,pra,reprint,twocolumn,amsmath,amssymb,citeautoscript]{revtex4-1}
\usepackage{graphicx}% Include figure files
\usepackage{dcolumn}% Align table columns on decimal point
\usepackage{bm}% bold math
\usepackage{latexsym,epsfig}
\usepackage{graphicx}
\usepackage{verbatim}
\usepackage{comment}
\usepackage{amsmath}
\usepackage{amssymb}
\usepackage{physics}
\usepackage{stmaryrd}
\usepackage{color}
\usepackage{epstopdf}
\usepackage{grffile}
\usepackage{lipsum}
\usepackage{enumitem}
\usepackage[dvipsnames]{xcolor}
\DeclareGraphicsExtensions{.eps}

\newcommand{\beq}{\begin{equation}}
\newcommand{\eeq}{\end{equation}}
\newcommand{\bea}{\begin{eqnarray}}
\newcommand{\eea}{\end{eqnarray}}
\newcommand{\ben}{\begin{eqnarray*}}
\newcommand{\een}{\end{eqnarray*}}
\newcommand{\bfig}{\begin{figure}}
\newcommand{\efig}{\end{figure}}
\usepackage{soul}

\usepackage{hyperref}
\hypersetup{
    colorlinks=true,      
    urlcolor=blue,
    citecolor=blue,
    linkcolor=blue
}

\begin{document}
%\title{Periodic and quasiperiodic extended Hatano-Nelson model} 
\title{Quasiperiodic and periodic extended Hatano-Nelson model: Anomalous complex-real transition and non-Hermitian skin effect} 
\author{Soumya Ranjan Padhi, Ashirbad Padhan, Sanchayan Banerjee and Tapan Mishra}

\email{mishratapan@gmail.com}

\affiliation{School of Physical Sciences, National Institute of Science Education and Research, Jatni 752050, India}

\affiliation{Homi Bhabha National Institute, Training School Complex, Anushaktinagar, Mumbai 400094, India}

\date{\today}

\begin{abstract}

We study the effect of quasiperiodic and periodic onsite potentials in a Hatano-Nelson model with next-nearest-neighbour hopping. By considering a non-reciprocal next-nearest-neighbour hopping and a quasiperiodic onsite potential under periodic boundary conditions, we show a breakdown of the typical correspondence between the delocalization-localization and complex-real transitions as a function of the potential strength. Moreover, we reveal that in the delocalized regime, when the potential strength increases, the eigenstates under open boundary conditions exhibit a bidirectional non-Hermitian skin effect, i.e., they tend to localize on both the edges instead of localizing on either of the edges. However, when a periodic onsite potential is considered, the system not only exhibits a bidirectional skin effect but also shows a complete direction reversal of the skin effect as a function of the onsite periodic potential. 

\end{abstract}

\maketitle

\section{Introduction} 
\label{sec:intro}

Non-Hermitian (NH) systems have triggered extensive research in the last couple of decades due to their unique properties which are different from their Hermitian counterparts. Unlike the Hermitian Hamiltonians which describe the dynamics of closed quantum systems with conserved probabilities and real eigenenergies, the non-Hermitian Hamiltonians often provide insights about various non-conservative systems including open quantum systems~\cite{rotter2017nonhermitian, AlvaroRamos, KawabataPRX}, solid-state systems with interaction~\cite{Yoshida,PhysRevLettShen, PhysRevLettYamamoto} and classical systems with gain and loss~\cite{acoustic, Acousticmetamaterials,metamaterials, NaturePhotonics}. 
%The self-adjoint nature of the Hamiltonian warrants a real eigenspectrum for the Hermitian case. 
Such non-Hermiticity in the Hamiltonian is known to reveal some striking features such as the existence of complex energyspectra, complex energy gaps, non-Hermitian skin effect (NHSE)~\cite{Yao2018, Yuce_2020, Sato2020, Li_2020, Zeng2022, Zhong2019, Okuma_2023, Roccati2021, Li_2021, Lin_2023}, exceptional points~\cite{Emil2021, Andre2019, meng2023, Nori2019, krasnok2021}, the failure of bulk boundary correspondence~\cite{Emil2018, Koch_2020, Cao2021, Xiao_2020, Jin2019}, non-trivial localization and spectral topology~\cite{longhi_2019, Liu_2020, Gandhi_2023, Datta_2024, Borgnia2020, Mandal_2024} etc. Exceptions have been found in the form of pseudo Hermiticity that gives rise to a completely real energyspectrum~\cite{Bender1998, Bender2007}. One such example can be a non-Hermitian system with balanced gain and loss satisfying parity-time ($\mathcal{P}\mathcal{T}$) symmetry. Although at the Hamiltonian level the $\mathcal{P}\mathcal{T}$ symmetry is respected, the eigenstates can still break this symmetry if the balance between gain and loss is broken, and the system exhibits complex eigenenergies. Thus, there occurs a transition between these so-called $\mathcal{P}\mathcal{T}$-unbroken and $\mathcal{P}\mathcal{T}$-broken phases when the non-Hermiticity increases in the system~\cite{Bender1998, Bender2007}.

One of the most fascinating aspects of non-Hermitian systems is the non-Hermitian skin effect, i.e., the accumulation of the eigenstates at a particular edge under open boundary conditions (OBCs). 
%This feature contradicts Bloch's prediction where the eigenstates are expected to be extended over all the lattice sites. 
The celebrated Hatano-Nelson (HN) model~\cite{Nelson_1996, Nelson_1997, Nelson1998} 
which describes a one-dimensional tight-binding chain with non-reciprocal nearest-neighbor hopping is a well known example of the NH systems that exhibits NHSE due to the dominant hopping along one direction compared to the other~\cite{Yao2018, Sato2020}. 
Due to the growing interest along this direction, the HN model has been widely explored and extended theoretically in the presence of long-range hopping, disorder and interaction~\cite{Ueda2019, Titus2022, Huang2020, Gong2020, Gil2021, Ichiro_2022, QiBoZeng2022}. On the experimental front, the emerging physics of this model has been realized using electric circuits~\cite{Liu2021, Zhang2023}, cold atoms in optical lattices~\cite{Dong, Zhou, Gong, BoYan}, acoustic systems~\cite{Zhang}, photonic lattices~\cite{Wang_2021}, single-photon quantum walk~\cite{Quan, PhysRevLettKunkun} and mechanical systems~\cite{mechanicalmetamaterial}.

On the other hand, the interplay between non-reciprocity and disorder in the HN model has been a topic of paramount interest in the context of non-Hermitian systems~\cite{longhi2021, Wei2021, cai2022, shu2021, Zeng2020, Zhou2023, Jia2023, Yuce2022, Peng2022, Taylor2021, Jia2021}. Studies have revealed that the HN model which breaks the $\mathcal{P}\mathcal{T}$ symmetry, undergoes a $\mathcal{P}\mathcal{T}$ symmetry broken-unbroken (or in other words, complex-real) phase transition as a function of disorder strength under periodic boundary conditions (PBCs). 
Efforts have also been made to understand the effect of quasiperiodic disorder in the HN model. Recent studies have revealed various interesting phenomena in the context of the quasiperiodic HN model such as localization transition, re-entrant localization, topological phases and NHSE etc~\cite{shuchen_2019, Rong_2020, Liu1_2021, Liu2_2021, Cai_2021, Gao_2021, Bin_2021, longhi_2021, Zhou_2021, Zhou_2022, Yong_2020, Zhai_2022}. While a great deal of research has been done in different variants of the HN model to understand the combined effect of non-reciprocal hopping and onsite potentials, the role of next-nearest-neighbour (NNN) non-reciprocal hopping in such systems has not been well investigated. 

In this work, we study the effect of quasiperiodic and periodic onsite potentials in an extended Hatano-Nelson (EHN) model, i.e., the HN model with non-reciprocal NNN hopping. We obtain that when a quasiperiodic potential is considered,  bulk localization of the eigenstates occurs when the strength of the potential is strong. The localization transition also gives rise to a complex-real transition. We show that although the localization of the eigenstates makes their eigenenergies real, the delocalization-localization transition and complex-real transition do not necessarily occur at the same critical point. Moreover, we obtain a
bidirectional NHSE for stronger quasiperiodic potential where the eigenstates occupy both the edges of the lattice under open boundary condition. Surprisingly, when a periodic potential is considered in place of the quasiperiodic potential, a complete direction reversal of the skin effect occurs with the increase in the potential strength for some fixed hopping strengths, i.e., the states localized in one end move to the other end.  

The rest of the paper is organized as follows. In Sec.~\ref{sec:model}, we present the model Hamiltonian. We explore the correspondence between the delocalization-localization and complex-real transitions in Sec.~\ref{sec:incom}, and discuss the results related to the NHSE for a quasiperiodic potential in Sec.~\ref{sec:NHSE}. Subsequently, for the case of periodic potential, we provide the results concerning the NHSE in Sec.~\ref{sec:com} and the wavepacket dynamics in Sec.~\ref{sec:dynamics}. Finally, in Sec.~\ref{sec:conc}, we summarize our results and provide a brief outlook.

\begin{figure}[t]
\centering
\includegraphics[width=0.7\columnwidth]{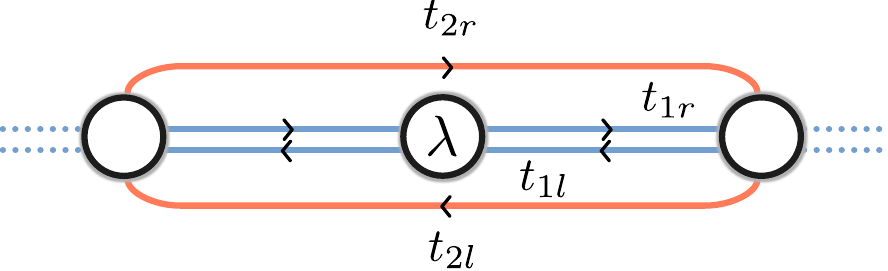}
\caption{Schematic of the lattice representing the EHN model with NN and NNN hopping strengths $t_{1l,r}$ and $t_{2l,r}$, respectively, and onsite potential strength $\lambda$. The arrows denote the direction of the hopping along the left or the right direction. }
\label{fig:fig1}
\end{figure}

\section{Model}
\label{sec:model}
The one-dimensional extended Hatano-Nelson model  with an onsite potential is represented by the Hamiltonian
\begin{align}
\label{eq:ham}
    H\ &= -\sum_{j} \big{(}t_{1l} c_j^\dagger c_{j+1} + t_{1r} c_{j+1}^\dagger c_j + t_{2l} c_j^\dagger c_{j+2} \\ \nonumber
    &+ t_{2r} c_{j+2}^\dagger c_j\big{)} 
    +  \lambda \sum_{j} \cos(2\pi\beta j) c_j^\dagger c_j,
\end{align} 
where $c_j^\dagger$ ($c_j$) is the creation (annihilation) operator of spinless fermions at the $j$th lattice site. The parameters $t_{1}$ and $t_{2}$ are the hopping amplitudes associated to the nearest-neighbour (NN) and next-nearest-neighbour (NNN) hopping strengths, respectively, and the subscript $l$ ($r$) denotes the direction of hopping towards left (right). $\lambda$ is the strength of the onsite potential and $\beta$ defines the nature of the potential. When $\beta$ is chosen to be a rational (an irrational) number, the potential is periodic (quasiperiodic) in nature. The incommensurate or quasiperiodic potential is achieved by setting $\beta=(\sqrt{5}-1)/2$ which is the inverse golden mean, and for the commensurate or periodic case, we assume $\beta=1/2$. We fix $t_{1l}=1$ for all our calculations which also sets the energy scale of the system.

Note that the Hamiltonian described above reduces to the well-known HN model when both $t_2$ and $\lambda$ are set to zero. It is well known that the HN model exhibits the localization of states at the left or right edge based on whether the hopping is dominant along the left or right direction, which is known as the NHSE. Here our objective is to investigate the characteristics of the NHSE in the presence of both quasiperiodic and periodic onsite potentials in the EHN model which we discuss in the following sections. First, we consider the case of onsite quasiperiodic potential and then the case of periodic potential.

\section{Effect of quasiperiodic potential}

In this section, we discuss the effect of the quasiperiodic potential on the EHN model. It has already been shown that the HN model with onsite quasiperiodic disorder exhibits a sharp delocalization-localization (DL) transition as a function of $\lambda$. Moreover, the spectrum undergoes a complex-real (CR) transition at a critical quasiperiodic disorder strength which coincides with the delocalization-localization transition~\cite{longhi_2019, shu2021}. In the following we investigate such DL and CR transitions by introducing the non-reciprocal NNN hopping. We first focus on the localization transition using systems with PBC that may arise due to the quasiperiodic potential and then investigate the NHSE using systems with OBC.

\subsection{Delocalization-localization and complex-real transitions}
\label{sec:incom}
To investigate the DL and CR  transitions for the system described by the EHN model shown in Eq.~(\ref{eq:ham}) we examine the properties of the  eigenstates and their corresponding eigenenergies obtained by exactly solving the Schr\"{o}dinger equation $H\psi_n=E_n\psi_n$. We choose $t_{1r}=0.6$ and $t_{2r}=0.4$ to study the competition between $t_{2l}$ and $\lambda$. The degree of localization of an eigenstate $|\psi_{n} \rangle$ can be quantified using the inverse participation ratio (IPR) defined as 
\begin{equation}
   \text{IPR}_n = \frac{\sum_{j=1}^{L} |\psi_{n,j}|^4} {(\langle \psi_{n}| \psi_{n} \rangle)^2}
\end{equation}
and the normalized participation ratio 
(NPR) defined as 
\begin{equation}
\text{NPR}_n = (L\times\text{IPR}_n)^{-1}.
\end{equation}
In the thermodynamic limit (i.e., for $L\to\infty$), IPR $=0$ and NPR $\neq0$ (IPR $\neq0$ and NPR $=0$) for a delocalized (localized) eigenstate. To obtain a complete information about the spectrum we compute the average IPR and NPR by summing over all the eigenstates which are defined as

\begin{align}
    \langle \text{IPR} \rangle = \frac{1}{L}\sum_{n=1}^{L}\text{IPR}_n
\end{align}
and
\begin{align}
    \langle \text{NPR} \rangle = \frac{1}{L}\sum_{n=1}^{L}\text{NPR}_n, 
\end{align}
respectively, where $\langle\cdot\rangle$ denotes the average over all states.

% \begin{align}
%     \eta = \text{log}_{10}(\langle \text{IPR} \rangle \times \langle \text{NPR} \rangle)
% \end{align}

\begin{figure}[t]
\centering
\includegraphics[width=1\columnwidth]{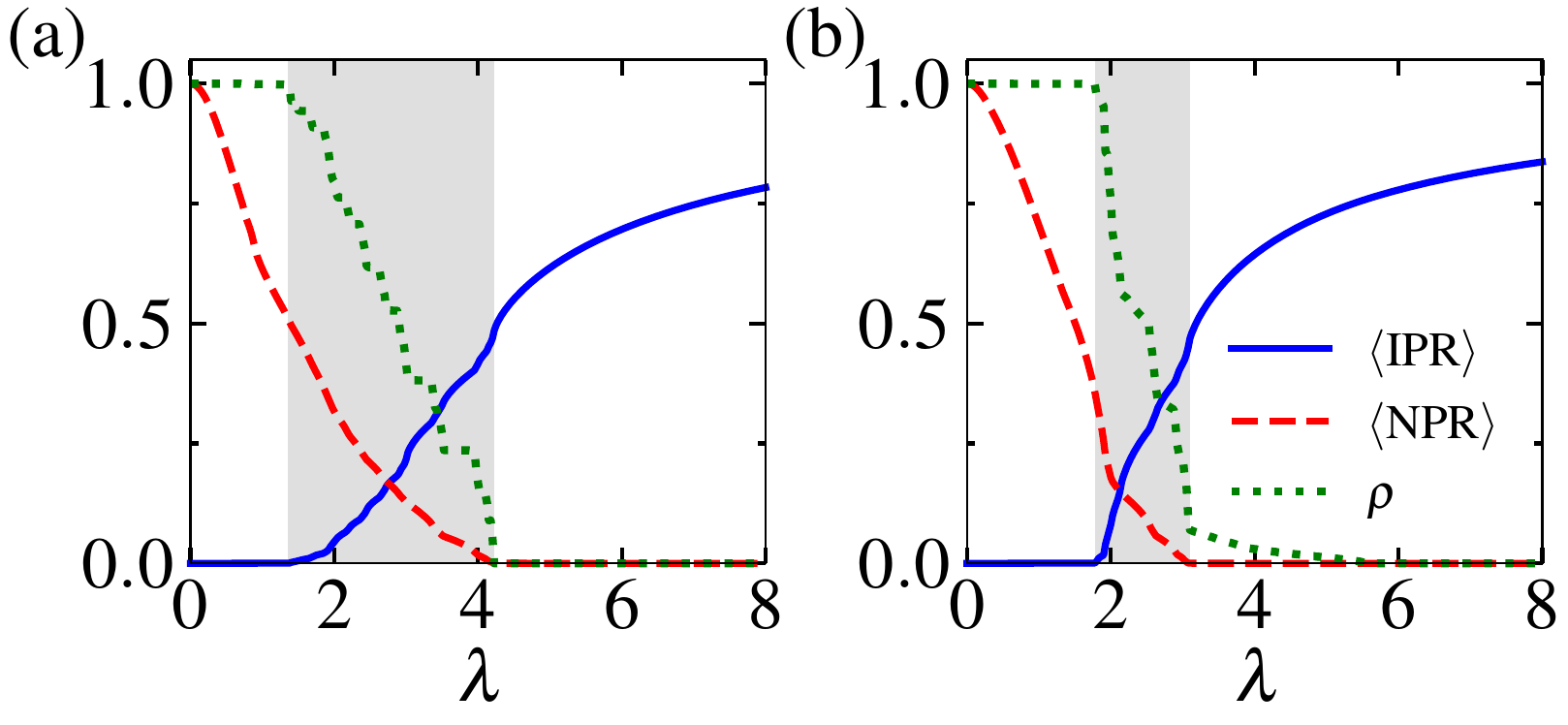}
\caption{$\langle \text{IPR} \rangle$ (blue solid line), $\langle \text{NPR} \rangle$ (red dashed line) and $\rho$ (green dotted line) are plotted as a function of $\lambda$ for $(t_{1r},t_{2r},t_{2l})=(0.6, 0.4, 0.5)$ in (a), and for $(t_{1r},t_{2r},t_{2l})=(0.6, 0.4, 0.1)$ in (b). Here a system of $L=6765$ lattice sites is considered with PBC. The gray regions mark the intermediate phase.}
\label{fig:fig3}
\end{figure}

\begin{figure}[t]
\centering
\includegraphics[width=1\columnwidth]{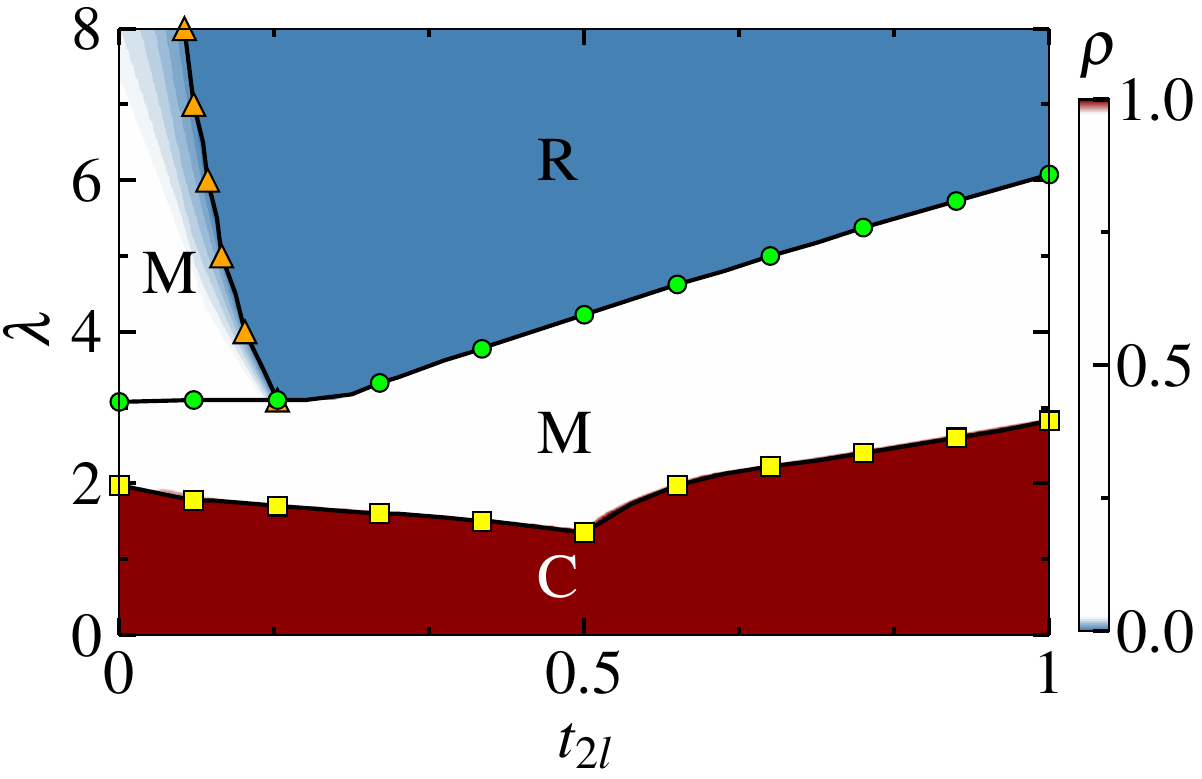}
\caption{Phase diagram in the $\lambda-t_{2l}$ plane obtained by plotting the density of states $\rho$ for $L=6765$ lattice sites. The red, white and blue regions depict the complex (C), mixed (M) and real (R) phases, respectively. The solid black lines with yellow squares and green circles mark the boundaries of the intermediate phase obtained through the extrapolated values of $\langle \text{IPR} \rangle$ and $\langle \text{NPR} \rangle$, respectively. In the region below (above) the intermediate region, the system is completely delocalized (localized). The solid black line with orange triangles marks the boundary between the mixed and real phases obtained through the extrapolated values of $\rho$. Here the extrapolation is performed with $L=610, 1597, 2584$ and $6765$ lattice sites.}
% \caption{Phase diagrams in the $\lambda-t_{2l}$ plane are shown as a function of $\eta$ in (a) to mark the localized (L), intermediate (I) and delocalized (D) phases, and as a function of $\text{max}|\zeta|$ in (b) to mark the complex (C) and real (R) phases. Here a system of $L=6765$ lattice sites is considered with periodic boundary condition.}
\label{fig:fig2}
\end{figure}

We plot $\langle \text{IPR} \rangle$ (blue solid line) and $\langle \text{NPR} \rangle$ (red dashed line) as a function of $\lambda$ in Fig.~\ref{fig:fig3} (a) for $t_{2l}=0.5$. This shows a transition from a delocalized phase to a localized phase via an intermediate phase between $\lambda\sim1.375$ and $\lambda\sim4.225$ where both the $\langle \text{IPR} \rangle$ and $\langle \text{NPR} \rangle$ are finite. We also find that for this choice of the parameters, the spectrum exhibits a complex-real transition which coincides with the delocalization-localization transition as has already been found in the standard HN model with disorder~\cite{Nelson_1996, Nelson_1997}. To quantify this feature we plot the density of states $\rho$ defined as
% \begin{align}
    % \text{max}|\zeta| = \text{max}_{n=1,...,L}(|\text{Im}(E_n)|)
% \end{align}
\begin{align}
    \rho = \frac{N}{L},
\end{align}
where $N$ is the number of imaginary eigenenergies as a function of $\lambda$ in Fig.~\ref{fig:fig3}(a) (green dotted line). This clearly shows that the complex-real transition of the spectrum occurs at the localization transition critical point, i.e., $\lambda\sim4.225$. However, we obtain a completely different scenario for smaller values of $t_{2l}$ where the complex-real transition does not coincide with the localization transition, rather occurs at higher values of $\lambda$. In Fig.~\ref{fig:fig3}(b) we plot $\langle \text{IPR} \rangle$, $\langle \text{NPR} \rangle$ and  $\rho$ as a function of $\lambda$ for $t_{2l}=0.1$. This clearly shows that the localization transition occurs at $\lambda\sim3.1$ and the complex-real transition occurs at a different value of quasiperiodic potential $\lambda\sim5.55$. This discrepancy also suggests that the localized spectrum for the range of $\lambda$ between $3.1$ and $5.55$ exhibits complex eigenenergies. Similar feature also persists for some other values of $t_{2l}$. To obtain a complete picture of the delocalization-localization and complex-real transitions, we plot $\rho$ as a function of $\lambda$ and $t_{2l}$ together with the localized and delocalized regions obtained from the extrapolated values of $\langle \text{IPR} \rangle$ and $\langle \text{NPR} \rangle$  in Fig.~\ref{fig:fig2}. The blue, white, and red regions denote the regimes where the spectrum is real (R), mixed (M) and complex (C), respectively. While the regions denoted by R (C) indicates completely real (complex) eigenenergies in the eigenspectrum, the region denoted by M exhibits a mixed spectrum of both real and complex eigenenergies. The region below (above) the black line with yellow squares (green circles) represents the delocalized (localized) region, and the region in between the two lines is the intermediate region. The mixed region at the top-left corner of Fig.~\ref{fig:fig2}(a) which is separated from the region-R by the orange triangles is the localized region with finite number of states possessing complex eigenenergies. It can be clearly seen that for $t_{2l}\gtrsim 0.17$, the transition to the localized phase coincides with the complex-real transition. However, for smaller values of $t_{2l}$, i.e. for $t_{2l} \lesssim 0.17$, localized states exhibit complex eigenenergies.

\begin{figure}[t]
\centering
\includegraphics[width=1\columnwidth]{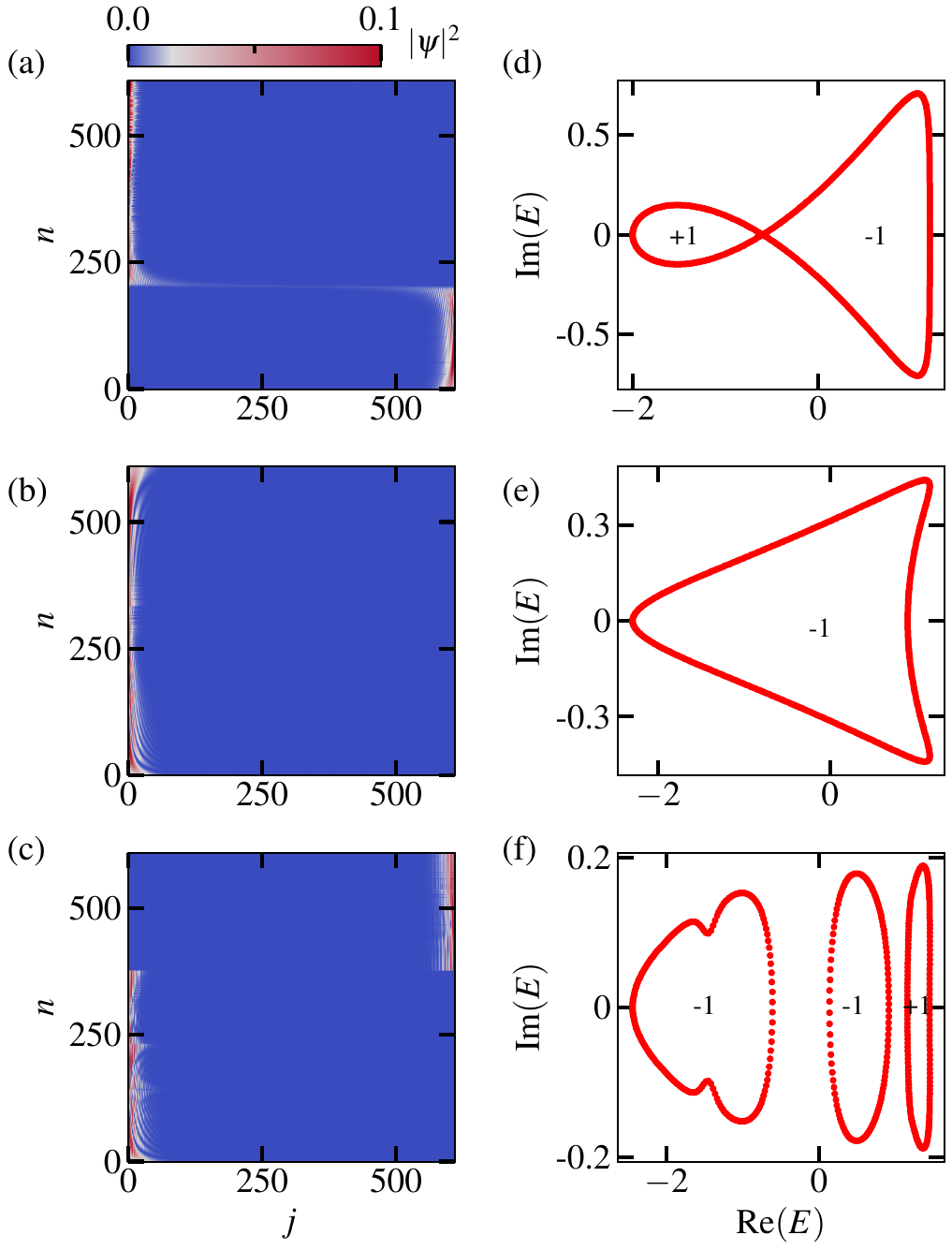}
\caption{Probability distribution of eigenstates ($|\psi|^2$) as a function of eigenstate index ($n$) and site index ($j$) for $(t_{2l}, \lambda)=(0.0, 0.0)$, $(t_{2l}, \lambda)=(0.3, 0.0)$ and $(t_{2l}, \lambda)=(0.3, 1.0)$ are plotted in (a), (b) and (c), respectively. The corresponding winding number ($w$) of the loops in the real vs imaginary eigenenergy plane is shown in (d), (e) and (f), respectively. Here we consider a system of size $L=610$. The color bar is scaled from 0 to 0.1 for clarity.}
\label{fig:fig4}
\end{figure}

\begin{figure}[t]
\centering
\includegraphics[width=1\columnwidth]{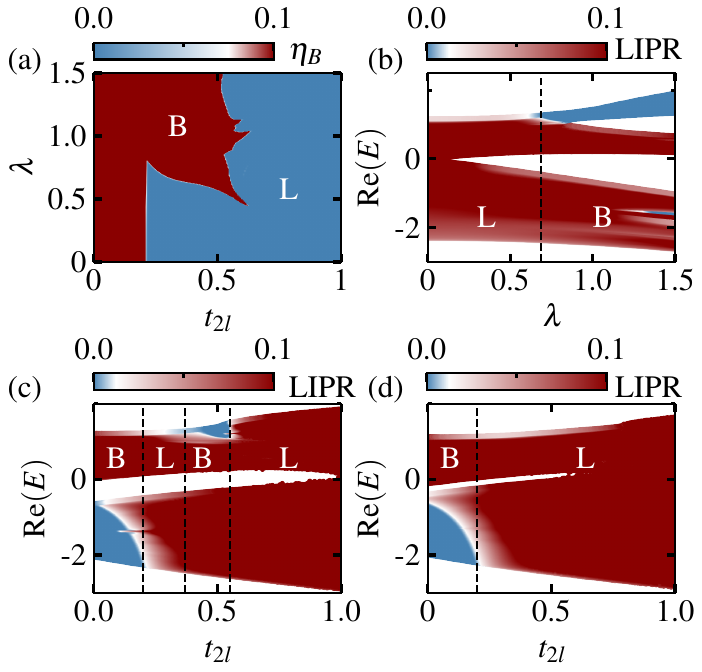}
\caption{(a) Phase diagram in $\lambda-t_{2l}$ plane is shown as a function of $\eta_B$. Left inverse participation ratios (LIPRs) are plotted as a function of real eigenenergies and $\lambda$ in (b) with $t_{2l}=0.3$. (c) and (d) LIPR as a function of the real eigenenergies and $t_{2l}$ for $\lambda=0.65$ and $0.25$, respectively. Here we consider $(t_{1r}, t_{2r})=(0.6, 0.4)$ and a system of length $L=610$ lattice sites under OBC.  The dashed vertical lines separate the regions possessing bidirectional and unidirectional NHSE. The region exhibiting bidirectional NHSE is denoted by B and the region for which the states are localized on the left edge of the lattice is denoted by L, i.e., the unidirectional NHSE. The color bars are scaled from 0 to 0.1 for clarity.}
\label{fig:fig5}
\end{figure}

\subsection{Non-Hermitian skin effect} 
\label{sec:NHSE}
In this subsection, we investigate the combined effect of the NNN hopping and quasiperiodic potential on the NHSE. Before turning on the quasiperiodic potential, we study the effect of the non-reciprocal NNN hopping on the NHSE of the HN model. To this end, we plot probabilities of all the eigenstates as a function of site index $j$ and eigenstate index $n$ for an open chain of size $L=610$ in Fig.~\ref{fig:fig4} for different values of $t_{2l}$ and $\lambda$ while keeping the other parameters as $t_{1r}=0.6$ and $t_{2r}=0.4$. We obtain that when both $t_{2l}$ and $\lambda$ are zero, the system exhibits the signature of bidirectional NHSE, i.e., both the edges of the chain are populated by some of the eigenstates (Fig.~\ref{fig:fig4}(a)). This is due to the combined effect of the right NN and NNN hoppings strengths. As $t_{2l}$ increases and becomes greater than $0.2$, the bidirectional signature vanishes and we get a usual unidirectional NHSE as shown in Fig.~\ref{fig:fig4}(b). This is due to the complete dominance of the left hopping on the system. At this point, we make $\lambda$ finite to study the behavior of the NHSE. For this purpose, we select  $t_{2l}= 0.3$ for which the system exhibits unidirectional NHSE and vary $\lambda$. 
Here, we find that after a particular value of the quasiperiodic potential i.e. $\lambda \sim 0.7$, the unidirectional NHSE becomes bidirectional in nature as shown in Fig.~\ref{fig:fig4}(c). To further quantify the bidirectional NHSE, we define the quantities called the left and right inverse participation ratios such as
\begin{align}
    \text{LIPR}_n = \frac{\sum_{j=1}^{L/2}|\psi_{n,j}|^4 }{(\langle \psi_{n}| \psi_{n} \rangle)^2} 
\end{align}
and
\begin{align}
     \text{RIPR}_n = \frac{\sum_{j=L/2+1}^{L}|\psi_{n,j}|^4}{(\langle \psi_{n}| \psi_{n} \rangle)^2},  
\end{align}
respectively. According to the definition, a right (left) localized eigenstate is characterised by LIPR $=0$ and RIPR $\neq 0$ (RIPR $=0$ and LIPR $\neq 0$). By taking the average of these quantities over all the eigenstates, we compute 
\begin{align}
     \text{$\eta_B$} = [\langle \text{RIPR} \rangle \times \langle \text{LIPR} \rangle] \times L^2 
\end{align}
% We can calculate the average of LIPR and RIPR over all the eigenstates, 
% \begin{align}
%      \text{$\langle \text{LIPR} \rangle$} = \frac{1}{L}\sum_n \text{LIPR}_n ,  
% \end{align}
% \begin{align}
%      \text{$\langle \text{RIPR} \rangle$} = \frac{1}{L}\sum_n \text{RIPR}_n ,  
% \end{align}
% Using these $\langle \text{LIPR} \rangle$ and $\langle \text{RIPR} \rangle$, we also plot a phase diagram of a quantity $\eta_B$, in the plane $\lambda$ vs $t_{2l}$, where 
such that when the system exhibits unidirectional (bidirectional) NHSE, the value of $\eta_B=0$ (finite). 

We plot $\eta_B$ as a function of $t_{2l}$ and $\lambda$ in Fig.~\ref{fig:fig5} (a) for $t_{1r}=0.6$ and $t_{2r}=0.4$. The red ($\eta_B\neq 0$) and blue ($\eta_B=0$) regions in the figure clearly depict the parameters for which the system exhibits bidirectional and unidirectional NHSE, respectively. To further clarify the nature of the NHSE we plot the LIPR for all the eigenstates with the corresponding eigenenergies as a function of $\lambda$ for $t_{2l}=0.3$ in Fig.~\ref{fig:fig5}(b). Finite values of LIPR for all the states up to $\lambda \sim 0.69$ suggests that the states are localized near the left edge of the lattice which is the unidirectional NHSE. However, for $\lambda \gtrsim 0.69$, some of the states exhibit LIPR$\sim 0$ (blue patches in Fig.~\ref{fig:fig5}(b)) and for the rest of the states LIPR is finite, indicating the bidirectional nature of the NHSE. Note that here we restrict ourselves to the value of $\lambda$ up to $\lambda\sim1.5$ since for higher values of $\lambda$, the eigenstates are localized at the bulk rather than at the edges and hence the notion of NHSE becomes irrelevant there (see Sec.~\ref{sec:incom}). 
It is important to note that, for a range of values of $\lambda$ between $0.49$ and $0.875$, we observe a re-entrant feature in the NHSE with an increase in $t_{2l}$, i.e., the initial bidirectional nature of the NHSE becomes unidirectional and then bidirectional and unidirectional again (see Fig.~\ref{fig:fig5}(a)). This re-entrant feature can be clearly seen by plotting LIPR as a function of the real part of the energy eigenenergies Re($E$) and $t_{2l}$ at $\lambda=0.65$, which is shown in Fig.~\ref{fig:fig5} (c). It can be seen that initially, for a range of $t_{2l}$, some states exhibit finite LIPR (red color) and some exhibits zero LIPR (blue color) indicating the bidirectional nature of the NHSE. As $t_{2l}$ increases, the entire spectrum becomes red which is the signature of the unidirectional NHSE. Further increase in $t_{2l}$ leads to the reappearance of both the blue and red regions together due to the reappearance of the bidirectional NHSE. Eventually the entire red region reappears indicating the unidirectional NHSE.  For comparison, we also show a case where such re-entrant behaviour of the NHSE is absent in Fig.~\ref{fig:fig5}(d) for $\lambda=0.25$.

Another striking feature that we obtain from the spectral behaviour, is the spectral topological invariant associated to the eigenenergies~\cite{Gong_2018}. Here we find that the eigenenergies corresponding to the spectrum exhibiting NHSE form loops in the complex energy plane which allows us to define a topological invariant. For our model, we define the topological invariant, i.e., the spectral winding number under PBC by the formula  
\begin{align}
    w=\lim_{L\to\infty} \frac{1}{2\pi i}  \int_{0}^{2\pi}d\theta\partial_{\theta}\log\big[\det\{H(\theta/L)-\varepsilon\}\big],
\label{eq:wind}
\end{align} 
where
\begin{align}
H(\theta/L) &= -\sum_{j} \big{(}t_{1l}e^{-i\theta/L} c_j^\dagger c_{j+1} + t_{1r}e^{i\theta/L} c_{j+1}^\dagger c_j \\ \nonumber
&+ t_{2l}e^{-2i\theta/L} c_j^\dagger c_{j+2} + t_{2r}e^{2i\theta/L} c_{j+2}^\dagger c_j\big{)} \\ \nonumber
&+  \lambda \sum_{j} \cos(2\pi\beta j) c_j^\dagger c_j \nonumber
\end{align}
is the modified Hamiltonian of the EHN model in the presence of a flux $\theta$ and $\varepsilon$ is a base energy. From intuition and the definition of $w$ we expect that when $\theta$ varies from $0$ to $2\pi$, the winding number will be finite only if a complex energy loop winds around $\varepsilon$ and it will vanish otherwise. We show that the winding presented here not only counts the number of times the base energy is wound by the eigenenergies but also provides significant information about the direction of the NHSE, i.e., the positive (negative) sign of $w$ corresponds to clockwise (anti-clockwise) winding of eigenenergies in complex energy plane, which indicates the localization of corresponding eigenstates to the right (left) edge of lattice. In Fig.~\ref{fig:fig4}(d) to (f), we plot $w$ for the eigenenergies for the parameters considered in Fig.~\ref{fig:fig4}(a) to (c) respectively. In Fig.~\ref{fig:fig4}(d), the twisted loop in the complex energy plane with $w = 1$ and $-1$ represents the localization of eigenstates at both edges, resulting in a bidirectional skin effect shown in Fig.~\ref{fig:fig4}(a). Also, it can be seen that when unidirectional skin effect appears, the complex energies form a single loop with $w= - 1$ (compare Fig.~\ref{fig:fig4}(b) with Fig.~\ref{fig:fig4}(e)). In Fig.~\ref{fig:fig4}(f), the left and middle loops yield $w = -1$, which signifies localization of corresponding eigenstates at the left edge and the right loop with $w = +1$, indicates localization at the right edge, which in turn forms a bidirectional skin effect shown in Fig.~\ref{fig:fig4}(c). 

From the above studies, it is understood that the quasiperiodic potential favours a bidirectional NHSE. Now we turn our attention to study the situation when the quasiperiodic potential is replaced by a periodic potential.

\section{Effect of periodic potential }
In this section, we explore the physics exhibited by the model shown in Eq.~\ref{eq:ham} in the presence of a periodic potential in place of the quasiperiodic potential. For this purpose, we set $\beta=1/2$ in Eq.~\ref{eq:ham}, which introduces a staggered onsite potential of strength $\lambda$ and $-\lambda$ at every alternate lattice site. In the following we first discuss the non-Hermitian skin effect and its signature in the wavepacket dynamics. 

\subsection{Non-Hermitian skin effect}
\label{sec:com}
\begin{figure}[t]
\centering
\includegraphics[width=1\columnwidth]{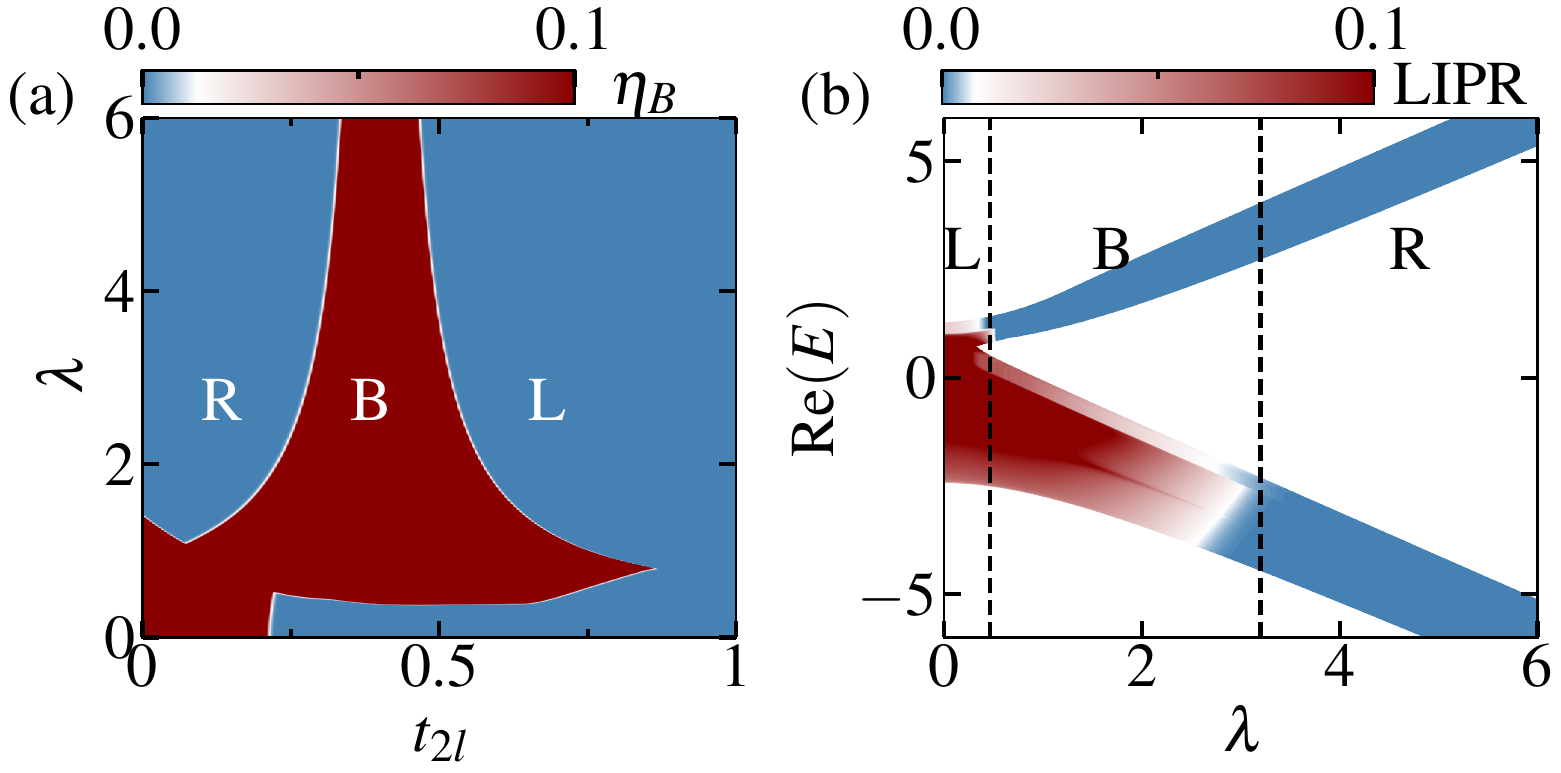}
\caption{(a) Phase diagram in the $\lambda-t_{2l}$ plane obtained through $\eta_B$. (b) Real eigenenergies as a function of $\lambda$ along with their corresponding left inverse participation ratio (LIPR) with fixed values of $(t_{1r},t_{2r},t_{2l})=(0.6, 0.4, 0.3)$ for a system size $L=610$. The dashed vertical lines separate the regions possessing bidirectional and unidirectional NHSE. The region exhibiting bidirectional NHSE is denoted by B and the region for which the states are localized on the left (right) edge of the lattice is denoted by L (R), i.e., the unidirectional NHSE. Here the color bars are scaled from 0 to 0.1 for clarity.}
\label{fig:fig6}
\end{figure}

Surprisingly, in the case of periodic onsite potential we find a completely different scenario which is depicted in the phase diagram shown in Fig.~\ref{fig:fig6}(a) obtained by plotting $\eta_B$ as a function of $\lambda$ and $t_{2l}$ for $t_{1r}=0.6$ and $t_{2r}=0.4$ which are same as the ones considered in Fig. 5(a). Contrary to the quasiperiodic case, we obtain that for vanishing and small values of $t_{2l}$ (i.e., between $0$ and $0.2$), the initial bidirectional nature of the NHSE (red region in Fig.~\ref{fig:fig6}(a)) changes to a unidirectional NHSE (blue region in Fig.~\ref{fig:fig6}(a)) as $\lambda$ increases. However, for values of  $t_{2l} \gtrsim 0.2$, we find that the system in the absence of $\lambda$ exhibits a unidirectional NHSE. In such a scenario, if $\lambda$ is turned on, we obtain surprising behaviour in the NHSE for a range of values of $t_{2l}$ where the initial unidirectional NHSE changes to bidirectional in nature and then becomes unidirectional again as a function of $\lambda$. The surprising nature of such transition is that while the unidirectional NHSE that appears for small values of $\lambda$ involves the localization of states on the left edge of the lattice, in the regime of large $\lambda$, they are localized on the right edge. This is quantified by plotting the LIPR for all the eigenstates along with the real energies and $\lambda$ at $t_{2l}=0.3$, in Fig.~\ref{fig:fig6}(b). It can be seen that initially, the LIPR of all the states is finite (red color) which indicates that the states are left localized. For  $0.47 \lesssim \lambda \lesssim 3.2$, we obtain that some of the states LIPR remain finite (red color) and for the rest of the states the LIPR vanishes (blue color), indicating the bidirectional nature of the skin effect. However, for $\lambda \gtrsim 3.2$, the LIPR of all the states vanishes (blue color), resulting in a complete right localization of the states. This behaviour of the NHSE from complete left localization to complete right localization through a region of bidirectional NHSE results in a complete direction reversal of the NHSE.

\begin{figure}[t]
\centering
\includegraphics[width=1\columnwidth]{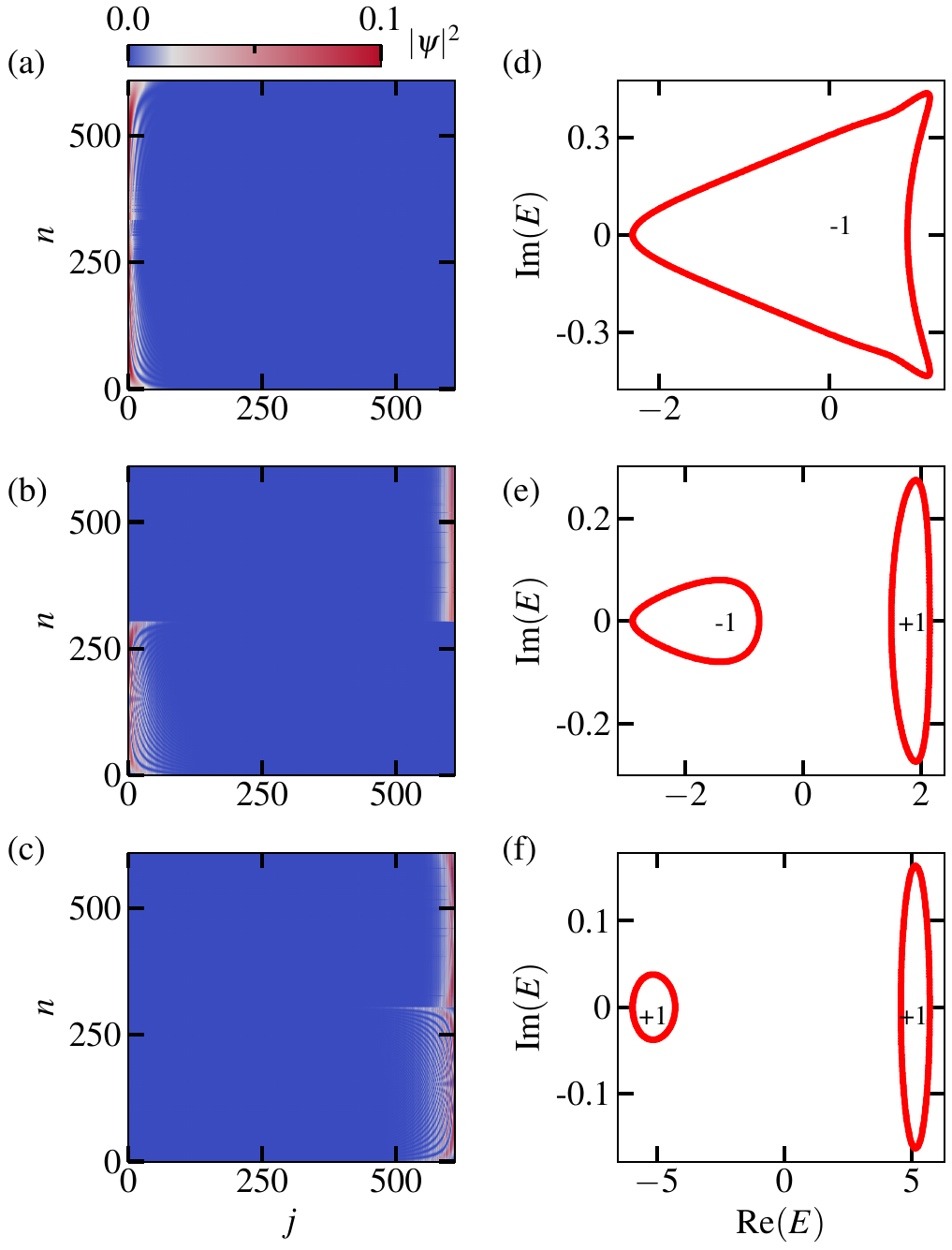}
\caption{Probability distribution of eigenstates ($|\psi|^2$) as a function of eigenstate index ($n$) and site index ($j$) for periodic potential strength $\lambda=0.15, 1.5$ and $5.0$ are plotted in (a), (b) and (c), respectively. The corresponding winding number ($w$) in the real vs imaginary eigenenergy plane is shown in (d), (e) and (f), respectively. Here we consider $(t_{1r}, t_{2r}, t_{2l}) = (0.6, 0.4, 0.3)$ and system size $L=610$. The color bar is scaled from 0 to 0.1 for clarity.}
\label{fig:fig7}
\end{figure}

To clearly visualize this NHSE, in Fig.~\ref{fig:fig7}(a) we plot the probability density $|\psi|^2$ as a function of the eigenstates and lattice sites for $\lambda=0.15$ which shows that all the eigenstates are localized near the left edge of the system. The corresponding real vs imaginary eigenenergies in PBC are shown in Fig.~\ref{fig:fig7}(d) which reveals the existence of a loop and if the winding number is computed for a base energy $\varepsilon$ lying inside the loop, it turns out to be $w=-1$. The negative sign here indicates an NHSE near the left edge. On the other hand, for $\lambda=1.5$ the states are localized at both the edges (see Fig.~\ref{fig:fig7}(b)). In Fig.~\ref{fig:fig7}(e), the left (right) loop yields $w=-1~(+1)$ for the winding around different real base energies. However, for $\lambda=5$, although the two loops are observed in the complex energy plane in Fig.~\ref{fig:fig7}(f), both the loops exhibit $w = +1$ and the behavior of probability densities in Fig.~\ref{fig:fig7}(c) indicates the complete localization of the states at the right edge of the lattice.   

To understand the behaviour of the spectral topology and its correspondence to the NHSE, we represent the concerned two-band model in the continuum space Hamiltonian
\begin{equation}
H_{k} =
\begin{bmatrix}
$$\lambda + t_{2r} e^{-ik} + t_{2l} e^{ik}$$  & $$t_{1r}  + t_{1l} e^{ik}$$ \\
$$t_{1l}  + t_{1r} e^{-ik}$$ &  $$ -\lambda + t_{2r} e^{-ik} + t_{2l} e^{ik}$$
\end{bmatrix},
\end{equation}
and the eigenenergies we get from $H_{k}$ are
\begin{equation}
    E_{\pm} = t_{2r} e^{-ik} + t_{2l} e^{ik} \pm \sqrt{\lambda^2 + 2 t_{1r} t_{1l} + t_{1r}^2 e^{-ik} + t_{1l}^2 e^{ik}}. 
    \label{eq:3d}
\end{equation}
\begin{figure}[t]
\centering
\includegraphics[width=1\columnwidth]{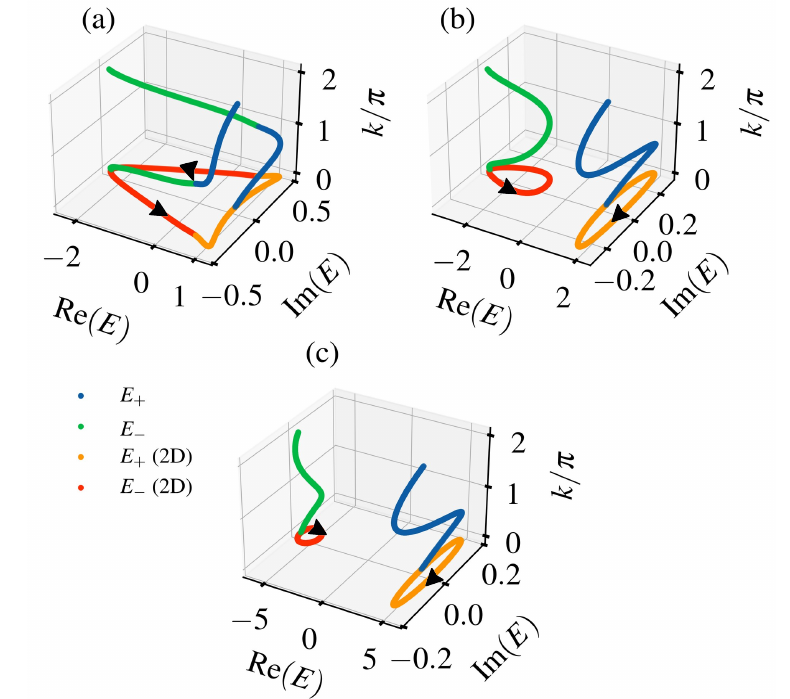}
\caption{Real vs imaginary eigenenergies as a function of $k$ at different values of (a) $\lambda$ = $0.15$, (b) $1.5$ and (c) $5.0$ with fixed values of $(t_{1r},t_{2r},t_{2l})=(0.6, 0.4, 0.3)$ are plotted using Eq.~\ref{eq:3d}. Their projections onto the 2D complex plane are also shown for clarity.}
\label{fig:fig8}
\end{figure} 
As discussed before, for the case of PBC  the formation of a loop in real vs imaginary energy plane corresponds to the NHSE and the direction of winding of this loop defines the direction of the localization of states in the case of OBC. This correspondence can be visualized by plotting the eigenenergies $E_{\pm}$ in the $x$- and $y$- axes as real and imaginary parts, respectively, for different values of $k$ (momentum) along the $z$-axis as shown in Fig.~\ref{fig:fig8}. As $k$ increases from $0$ to $2 \pi$, the nature of eigenenergy that evolves in the three-dimensional space tells the direction of the loops formed in their projections onto the two-dimensional (2D) complex plane ($x$-$y$ plane) which also discloses the direction of the NHSE. It can be seen that the real vs imaginary energy diagram obtained this way matches well with the numerical results shown in Fig.~\ref{fig:fig7} (right panel).

The complete direction reversal in skin effect can be attributed to the onsite staggered potential and the interplay between the non-reciprocal NN and NNN hopping strengths which are $t_{1l}=1$, $t_{1r}=0.6$, $t_{2l}=0.3$ and $t_{2r}=0.4$. In the absence of the onsite potential, i.e., $\lambda=0$, since the effective left hopping dominates over the right hopping, we get the NHSE where the states are completely localized along the left edge of the lattice. Now as the value of $\lambda$ increases, the tunneling probability towards the left reduces and at the same time increases along the right due to the staggered nature of the onsite potential. Consequently, as $\lambda$ exceeds $ 0.47$, some eigenstates begin to localize at the right edge, while others continue to remain localized at the left edge, which effectively gives rise to a bidirectional NHSE. Furthermore, as the potential continues to increase, especially after reaching  $\lambda\gtrsim3.2$,  all the eigenstates prefer to shift towards the right edge of the system, constituting a full direction reversal. In the quasiperiodic scenario also, with an increase in the potential strength, we witness that some eigenstates start to localize at the right edge, effectively inducing a bidirectional skin effect, contrasting to the case when quasiperiodic potential is zero, where all the states were localized at the left. However, we do not observe a complete directional reversal in this case, as there always remains some finite tunneling probability to the left owing to the incommensurate nature of the lattice potential. Instead, as the quasiperiodic potential approaches the critical threshold, a prominent onset of bulk localization becomes evident (not shown).

\subsection{Wavepacket dynamics}
\label{sec:dynamics}
\begin{figure}[t]
\centering
\includegraphics[width=1\columnwidth]{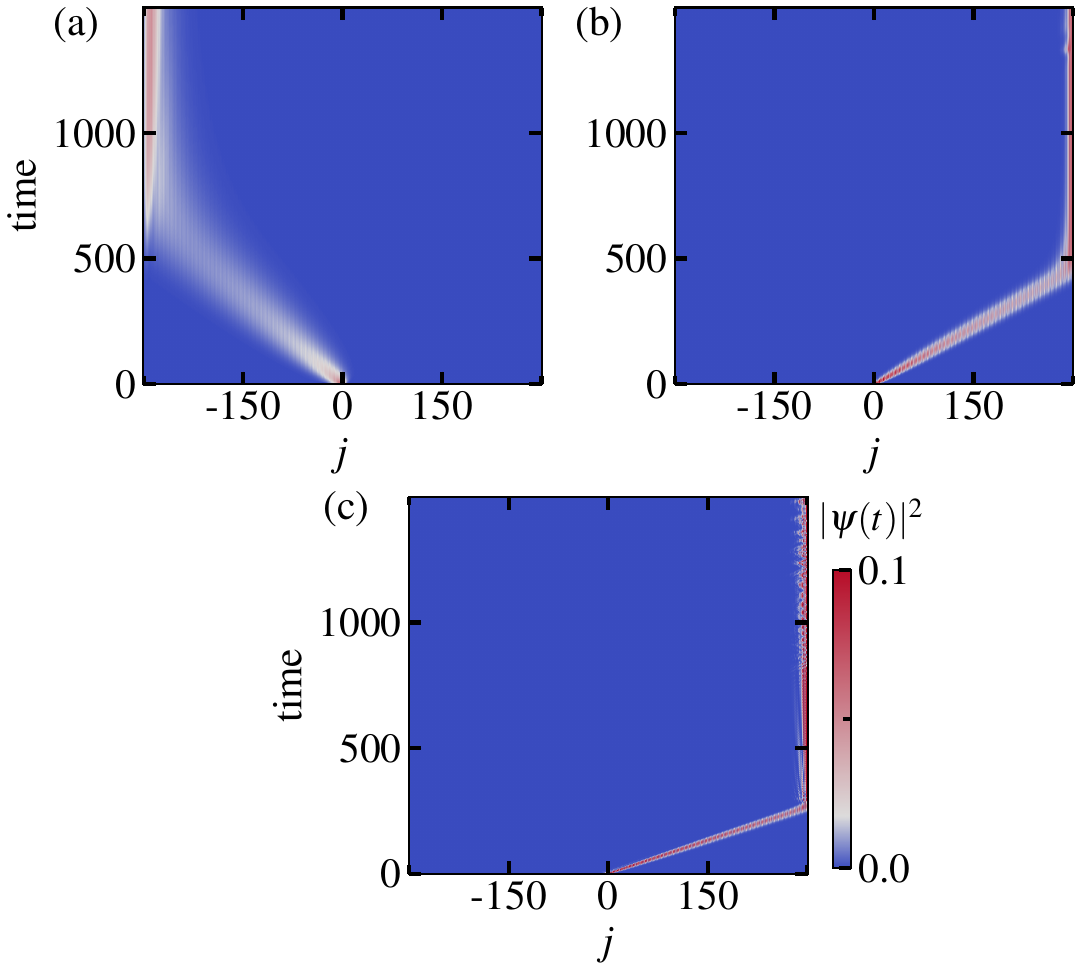}

\caption{Probability density ($|\psi(t)|^2$) as a function of time and site index ($j$) with $(t_{1r}, t_{2r}, t_{2l})=(0.6, 0.4, 0.3)$ considering periodic potential strength (a) $\lambda=0.15$, (b) $1.5$ and (c) $5.0$ for a system size $L=601$. Here the color bar is scaled from 0 to 0.1 for clarity.}
\label{fig:fig9}
\end{figure} 

In this subsection we obtain the signature of the complete direction reversal of the NHSE using in the dynamics of an initial wavepacket under OBC. The time evolved state at time $t$ from an initial state $\ket{\psi(0)}$ can be obtained by solving the time dependent Schr\"{o}dinger equation as $|\tilde{\psi}(t)\rangle=e^{-iHt}\ket{\psi(0)}$. However, since the evolution turns out to be non-unitary for a non-Hermitian system, $|\tilde{\psi}(t)\rangle$ is not normalized~\cite{longhi2021, Ichiro_2022}. Thus, we first normalize the state as  $\ket{\psi(t)}=|\tilde{\psi}(t)\rangle/\sqrt{\langle \tilde{\psi}(t)|\tilde{\psi}(t)\rangle}$ and study the  dynamics. For this purpose we choose an initial state describing a particle localized at the central lattice site. In Fig.~\ref{fig:fig9}, we plot the probability density $|\psi(t)|^2$ as a function of time and site indices for $(t_{1r},t_{2r},t_{2l})=(0.6, 0.4, 0.3)$, $\lambda=0.15, 1.5$ and $5$. For $\lambda=0.15$, the centrally localized wavepacket at $t=0$ gradually evolves towards the left and eventually becomes localized at the left edge (see Fig.~\ref{fig:fig9}(a)) indicating the unidirectional NHSE for which the states are localized on the left edge.  Interestingly for $\lambda=5$, the wavepacket moves towards the opposite direction and becomes localized at the right edge (see Fig.~\ref{fig:fig9}(c)). This is a dynamical signature of the complete directional reversal of the NHSE. 
However, in the case of $\lambda=1.5$, although a bidirectional skin effect is observed in the static case, we find that the localization of the particle in the dynamics is  directed towards the right edge. The reason behind this strange phenomenon lies in the basics of the NHSE. While performing non-unitary time evolution of a non-Hermitian system with complex energies, the time evolution operator $U=e^{-iHt}$ exhibits either a growing or decaying effect on the wavepacket due to the complex nature of the energy, where corresponding eigenenergies of Hamiltonian are $E=\text{Re}(E)-i\text{Im}(E)$ or $\text{Re}(E)+i\text{Im}(E)$. This phenomenon gives rise to the accumulation of all energy modes at a particular edge, known as the non-Hermitian skin effect (NHSE). 
In Fig.~\ref{fig:fig7}(e), for $\lambda=1.5$, two loops are formed in the complex plane of eigenenergy. The maximum imaginary energy of the loop with positive real energy (the right loop) with winding number $+1$ is significantly greater than the other loop with winding numbers $-1$ (the left loop). Consequently, when the unitary operator prepared from the Hamiltonian of the system acts on the initial wavepacket, the loop with the greater maximum energy dominates to exhibit the skin effect over the other loop. Thus, we observe the skin effect directed towards the right, which is associated with the winding number $+1$ of the loop with positive real energy (the right one).

%----------------------------------------------------------%
%----------------------------------------------------------%
\section{Conclusion}
\label{sec:conc}
In summary, our study focused on investigating the impact of next-nearest-neighbor hopping on the characteristic features of the HN model in the presence of an onsite potential. When the potential is quasiperiodic in nature, we obtain a delocalized-localized phase transition occurring at a different critical point compared to the complex-real transition as a function of the potential. Furthermore, we showed that in the delocalized phase, the NHSE originally directed towards a single direction, turns bidirectional in nature when the potential is tuned. Then we demonstrated the effect of a periodic potential instead of the quasiperiodic potential and revealed that not only the system exhibits a bidirectional NHSE but also a complete direction reversal of the skin effect when the strength of the potential increases. We demonstrated this feature of complete direction reversal of the skin effect from the wavepacket dynamics. 

Our study is focused on a model which is a simple extension of the well known HN model where the interplay between the NN and NNN hopping along with the onsite potential reveals non-trivial features. We predict the complete direction reversal phenomenon in an HN model by tuning the onsite potential of the simplest possible form and hence the model considered here can in principle be simulated in the state-of-the-art experiments. The system considered here can also be extended to study the effect of interaction on the NHSE. The complete reversal of NHSE can also be utilized as a switch to control the localization of eigenstates at either of the edges.

\section{Acknowledgement}
T.M. acknowledges support from Science and Engineering Research Board (SERB), Govt. of India, through project No. MTR/2022/000382 and STR/2022/000023.

\bibliography{ref}

\begin{thebibliography}{83}%
\makeatletter
\providecommand \@ifxundefined [1]{%
 \@ifx{#1\undefined}
}%
\providecommand \@ifnum [1]{%
 \ifnum #1\expandafter \@firstoftwo
 \else \expandafter \@secondoftwo
 \fi
}%
\providecommand \@ifx [1]{%
 \ifx #1\expandafter \@firstoftwo
 \else \expandafter \@secondoftwo
 \fi
}%
\providecommand \natexlab [1]{#1}%
\providecommand \enquote  [1]{``#1''}%
\providecommand \bibnamefont  [1]{#1}%
\providecommand \bibfnamefont [1]{#1}%
\providecommand \citenamefont [1]{#1}%
\providecommand \href@noop [0]{\@secondoftwo}%
\providecommand \href [0]{\begingroup \@sanitize@url \@href}%
\providecommand \@href[1]{\@@startlink{#1}\@@href}%
\providecommand \@@href[1]{\endgroup#1\@@endlink}%
\providecommand \@sanitize@url [0]{\catcode `\\12\catcode `\$12\catcode `\&12\catcode `\#12\catcode `\^12\catcode `\_12\catcode `\%12\relax}%
\providecommand \@@startlink[1]{}%
\providecommand \@@endlink[0]{}%
\providecommand \url  [0]{\begingroup\@sanitize@url \@url }%
\providecommand \@url [1]{\endgroup\@href {#1}{\urlprefix }}%
\providecommand \urlprefix  [0]{URL }%
\providecommand \Eprint [0]{\href }%
\providecommand \doibase [0]{http://dx.doi.org/}%
\providecommand \selectlanguage [0]{\@gobble}%
\providecommand \bibinfo  [0]{\@secondoftwo}%
\providecommand \bibfield  [0]{\@secondoftwo}%
\providecommand \translation [1]{[#1]}%
\providecommand \BibitemOpen [0]{}%
\providecommand \bibitemStop [0]{}%
\providecommand \bibitemNoStop [0]{.\EOS\space}%
\providecommand \EOS [0]{\spacefactor3000\relax}%
\providecommand \BibitemShut  [1]{\csname bibitem#1\endcsname}%
\let\auto@bib@innerbib\@empty
%</preamble>
\bibitem [{\citenamefont {Rotter}(2017)}]{rotter2017nonhermitian}%
  \BibitemOpen
  \bibfield  {author} {\bibinfo {author} {\bibfnamefont {I.}~\bibnamefont {Rotter}},\ }\href@noop {} {} (\bibinfo {year} {2017}),\ \Eprint {http://arxiv.org/abs/1707.03298} {arXiv:1707.03298} \BibitemShut {NoStop}%
\bibitem [{\citenamefont {G\'omez-Le\'on}\ \emph {et~al.}(2022)\citenamefont {G\'omez-Le\'on}, \citenamefont {Ramos}, \citenamefont {Gonz\'alez-Tudela},\ and\ \citenamefont {Porras}}]{AlvaroRamos}%
  \BibitemOpen
  \bibfield  {author} {\bibinfo {author} {\bibfnamefont {A.}~\bibnamefont {G\'omez-Le\'on}}, \bibinfo {author} {\bibfnamefont {T.}~\bibnamefont {Ramos}}, \bibinfo {author} {\bibfnamefont {A.}~\bibnamefont {Gonz\'alez-Tudela}}, \ and\ \bibinfo {author} {\bibfnamefont {D.}~\bibnamefont {Porras}},\ }\href {\doibase 10.1103/PhysRevA.106.L011501} {\bibfield  {journal} {\bibinfo  {journal} {Phys. Rev. A}\ }\textbf {\bibinfo {volume} {106}},\ \bibinfo {pages} {L011501} (\bibinfo {year} {2022})}\BibitemShut {NoStop}%
\bibitem [{\citenamefont {Kawabata}\ \emph {et~al.}(2023)\citenamefont {Kawabata}, \citenamefont {Xiao}, \citenamefont {Ohtsuki},\ and\ \citenamefont {Shindou}}]{KawabataPRX}%
  \BibitemOpen
  \bibfield  {author} {\bibinfo {author} {\bibfnamefont {K.}~\bibnamefont {Kawabata}}, \bibinfo {author} {\bibfnamefont {Z.}~\bibnamefont {Xiao}}, \bibinfo {author} {\bibfnamefont {T.}~\bibnamefont {Ohtsuki}}, \ and\ \bibinfo {author} {\bibfnamefont {R.}~\bibnamefont {Shindou}},\ }\href {\doibase 10.1103/PRXQuantum.4.040312} {\bibfield  {journal} {\bibinfo  {journal} {PRX Quantum}\ }\textbf {\bibinfo {volume} {4}},\ \bibinfo {pages} {040312} (\bibinfo {year} {2023})}\BibitemShut {NoStop}%
\bibitem [{\citenamefont {Yoshida}\ \emph {et~al.}(2018)\citenamefont {Yoshida}, \citenamefont {Peters},\ and\ \citenamefont {Kawakami}}]{Yoshida}%
  \BibitemOpen
  \bibfield  {author} {\bibinfo {author} {\bibfnamefont {T.}~\bibnamefont {Yoshida}}, \bibinfo {author} {\bibfnamefont {R.}~\bibnamefont {Peters}}, \ and\ \bibinfo {author} {\bibfnamefont {N.}~\bibnamefont {Kawakami}},\ }\href {\doibase 10.1103/PhysRevB.98.035141} {\bibfield  {journal} {\bibinfo  {journal} {Phys. Rev. B}\ }\textbf {\bibinfo {volume} {98}},\ \bibinfo {pages} {035141} (\bibinfo {year} {2018})}\BibitemShut {NoStop}%
\bibitem [{\citenamefont {Shen}\ and\ \citenamefont {Fu}(2018)}]{PhysRevLettShen}%
  \BibitemOpen
  \bibfield  {author} {\bibinfo {author} {\bibfnamefont {H.}~\bibnamefont {Shen}}\ and\ \bibinfo {author} {\bibfnamefont {L.}~\bibnamefont {Fu}},\ }\href {\doibase 10.1103/PhysRevLett.121.026403} {\bibfield  {journal} {\bibinfo  {journal} {Phys. Rev. Lett.}\ }\textbf {\bibinfo {volume} {121}},\ \bibinfo {pages} {026403} (\bibinfo {year} {2018})}\BibitemShut {NoStop}%
\bibitem [{\citenamefont {Yamamoto}\ \emph {et~al.}(2019)\citenamefont {Yamamoto}, \citenamefont {Nakagawa}, \citenamefont {Adachi}, \citenamefont {Takasan}, \citenamefont {Ueda},\ and\ \citenamefont {Kawakami}}]{PhysRevLettYamamoto}%
  \BibitemOpen
  \bibfield  {author} {\bibinfo {author} {\bibfnamefont {K.}~\bibnamefont {Yamamoto}}, \bibinfo {author} {\bibfnamefont {M.}~\bibnamefont {Nakagawa}}, \bibinfo {author} {\bibfnamefont {K.}~\bibnamefont {Adachi}}, \bibinfo {author} {\bibfnamefont {K.}~\bibnamefont {Takasan}}, \bibinfo {author} {\bibfnamefont {M.}~\bibnamefont {Ueda}}, \ and\ \bibinfo {author} {\bibfnamefont {N.}~\bibnamefont {Kawakami}},\ }\href {\doibase 10.1103/PhysRevLett.123.123601} {\bibfield  {journal} {\bibinfo  {journal} {Phys. Rev. Lett.}\ }\textbf {\bibinfo {volume} {123}},\ \bibinfo {pages} {123601} (\bibinfo {year} {2019})}\BibitemShut {NoStop}%
\bibitem [{\citenamefont {{Cummer}}\ \emph {et~al.}(2016)\citenamefont {{Cummer}}, \citenamefont {{Christensen}},\ and\ \citenamefont {{Al{\`u}}}}]{acoustic}%
  \BibitemOpen
  \bibfield  {author} {\bibinfo {author} {\bibfnamefont {S.~A.}\ \bibnamefont {{Cummer}}}, \bibinfo {author} {\bibfnamefont {J.}~\bibnamefont {{Christensen}}}, \ and\ \bibinfo {author} {\bibfnamefont {A.}~\bibnamefont {{Al{\`u}}}},\ }\href {\doibase 10.1038/natrevmats.2016.1} {\bibfield  {journal} {\bibinfo  {journal} {Nature Reviews Materials}\ }\textbf {\bibinfo {volume} {1}},\ \bibinfo {pages} {16001} (\bibinfo {year} {2016})}\BibitemShut {NoStop}%
\bibitem [{\citenamefont {Ma}\ and\ \citenamefont {Sheng}(2016)}]{Acousticmetamaterials}%
  \BibitemOpen
  \bibfield  {author} {\bibinfo {author} {\bibfnamefont {G.}~\bibnamefont {Ma}}\ and\ \bibinfo {author} {\bibfnamefont {P.}~\bibnamefont {Sheng}},\ }\href {\doibase 10.1126/sciadv.1501595} {\bibfield  {journal} {\bibinfo  {journal} {Science Advances}\ }\textbf {\bibinfo {volume} {2}},\ \bibinfo {pages} {e1501595} (\bibinfo {year} {2016})}\BibitemShut {NoStop}%
\bibitem [{\citenamefont {Zangeneh-Nejad}\ and\ \citenamefont {Fleury}(2019)}]{metamaterials}%
  \BibitemOpen
  \bibfield  {author} {\bibinfo {author} {\bibfnamefont {F.}~\bibnamefont {Zangeneh-Nejad}}\ and\ \bibinfo {author} {\bibfnamefont {R.}~\bibnamefont {Fleury}},\ }\href {\doibase https://doi.org/10.1016/j.revip.2019.100031} {\bibfield  {journal} {\bibinfo  {journal} {Reviews in Physics}\ }\textbf {\bibinfo {volume} {4}},\ \bibinfo {pages} {100031} (\bibinfo {year} {2019})}\BibitemShut {NoStop}%
\bibitem [{\citenamefont {{Feng}}\ \emph {et~al.}(2017)\citenamefont {{Feng}}, \citenamefont {{El-Ganainy}},\ and\ \citenamefont {{Ge}}}]{NaturePhotonics}%
  \BibitemOpen
  \bibfield  {author} {\bibinfo {author} {\bibfnamefont {L.}~\bibnamefont {{Feng}}}, \bibinfo {author} {\bibfnamefont {R.}~\bibnamefont {{El-Ganainy}}}, \ and\ \bibinfo {author} {\bibfnamefont {L.}~\bibnamefont {{Ge}}},\ }\href {\doibase 10.1038/s41566-017-0031-1} {\bibfield  {journal} {\bibinfo  {journal} {Nature Photonics}\ }\textbf {\bibinfo {volume} {11}},\ \bibinfo {pages} {752} (\bibinfo {year} {2017})}\BibitemShut {NoStop}%
\bibitem [{\citenamefont {Yao}\ and\ \citenamefont {Wang}(2018)}]{Yao2018}%
  \BibitemOpen
  \bibfield  {author} {\bibinfo {author} {\bibfnamefont {S.}~\bibnamefont {Yao}}\ and\ \bibinfo {author} {\bibfnamefont {Z.}~\bibnamefont {Wang}},\ }\href {\doibase 10.1103/PhysRevLett.121.086803} {\bibfield  {journal} {\bibinfo  {journal} {Phys. Rev. Lett.}\ }\textbf {\bibinfo {volume} {121}},\ \bibinfo {pages} {086803} (\bibinfo {year} {2018})}\BibitemShut {NoStop}%
\bibitem [{\citenamefont {Yuce}(2020)}]{Yuce_2020}%
  \BibitemOpen
  \bibfield  {author} {\bibinfo {author} {\bibfnamefont {C.}~\bibnamefont {Yuce}},\ }\href {\doibase 10.1016/j.physleta.2019.126094} {\bibfield  {journal} {\bibinfo  {journal} {Physics Letters A}\ }\textbf {\bibinfo {volume} {384}},\ \bibinfo {pages} {126094} (\bibinfo {year} {2020})}\BibitemShut {NoStop}%
\bibitem [{\citenamefont {Okuma}\ \emph {et~al.}(2020)\citenamefont {Okuma}, \citenamefont {Kawabata}, \citenamefont {Shiozaki},\ and\ \citenamefont {Sato}}]{Sato2020}%
  \BibitemOpen
  \bibfield  {author} {\bibinfo {author} {\bibfnamefont {N.}~\bibnamefont {Okuma}}, \bibinfo {author} {\bibfnamefont {K.}~\bibnamefont {Kawabata}}, \bibinfo {author} {\bibfnamefont {K.}~\bibnamefont {Shiozaki}}, \ and\ \bibinfo {author} {\bibfnamefont {M.}~\bibnamefont {Sato}},\ }\href {\doibase 10.1103/PhysRevLett.124.086801} {\bibfield  {journal} {\bibinfo  {journal} {Phys. Rev. Lett.}\ }\textbf {\bibinfo {volume} {124}},\ \bibinfo {pages} {086801} (\bibinfo {year} {2020})}\BibitemShut {NoStop}%
\bibitem [{\citenamefont {Li}\ \emph {et~al.}(2020{\natexlab{a}})\citenamefont {Li}, \citenamefont {Lee}, \citenamefont {Mu},\ and\ \citenamefont {Gong}}]{Li_2020}%
  \BibitemOpen
  \bibfield  {author} {\bibinfo {author} {\bibfnamefont {L.}~\bibnamefont {Li}}, \bibinfo {author} {\bibfnamefont {C.~H.}\ \bibnamefont {Lee}}, \bibinfo {author} {\bibfnamefont {S.}~\bibnamefont {Mu}}, \ and\ \bibinfo {author} {\bibfnamefont {J.}~\bibnamefont {Gong}},\ }\href {\doibase 10.1038/s41467-020-18917-4} {\bibfield  {journal} {\bibinfo  {journal} {Nature Communications}\ }\textbf {\bibinfo {volume} {11}},\ \bibinfo {pages} {5491} (\bibinfo {year} {2020}{\natexlab{a}})}\BibitemShut {NoStop}%
\bibitem [{\citenamefont {Zeng}(2022{\natexlab{a}})}]{Zeng2022}%
  \BibitemOpen
  \bibfield  {author} {\bibinfo {author} {\bibfnamefont {Q.-B.}\ \bibnamefont {Zeng}},\ }\href {\doibase 10.1103/PhysRevB.106.235411} {\bibfield  {journal} {\bibinfo  {journal} {Phys. Rev. B}\ }\textbf {\bibinfo {volume} {106}},\ \bibinfo {pages} {235411} (\bibinfo {year} {2022}{\natexlab{a}})}\BibitemShut {NoStop}%
\bibitem [{\citenamefont {Song}\ \emph {et~al.}(2019)\citenamefont {Song}, \citenamefont {Yao},\ and\ \citenamefont {Wang}}]{Zhong2019}%
  \BibitemOpen
  \bibfield  {author} {\bibinfo {author} {\bibfnamefont {F.}~\bibnamefont {Song}}, \bibinfo {author} {\bibfnamefont {S.}~\bibnamefont {Yao}}, \ and\ \bibinfo {author} {\bibfnamefont {Z.}~\bibnamefont {Wang}},\ }\href {\doibase 10.1103/PhysRevLett.123.170401} {\bibfield  {journal} {\bibinfo  {journal} {Phys. Rev. Lett.}\ }\textbf {\bibinfo {volume} {123}},\ \bibinfo {pages} {170401} (\bibinfo {year} {2019})}\BibitemShut {NoStop}%
\bibitem [{\citenamefont {Okuma}\ and\ \citenamefont {Sato}(2023)}]{Okuma_2023}%
  \BibitemOpen
  \bibfield  {author} {\bibinfo {author} {\bibfnamefont {N.}~\bibnamefont {Okuma}}\ and\ \bibinfo {author} {\bibfnamefont {M.}~\bibnamefont {Sato}},\ }\href {\doibase 10.1146/annurev-conmatphys-040521-033133} {\bibfield  {journal} {\bibinfo  {journal} {Annual Review of Condensed Matter Physics}\ }\textbf {\bibinfo {volume} {14}},\ \bibinfo {pages} {83–107} (\bibinfo {year} {2023})}\BibitemShut {NoStop}%
\bibitem [{\citenamefont {Roccati}(2021)}]{Roccati2021}%
  \BibitemOpen
  \bibfield  {author} {\bibinfo {author} {\bibfnamefont {F.}~\bibnamefont {Roccati}},\ }\href {\doibase 10.1103/PhysRevA.104.022215} {\bibfield  {journal} {\bibinfo  {journal} {Phys. Rev. A}\ }\textbf {\bibinfo {volume} {104}},\ \bibinfo {pages} {022215} (\bibinfo {year} {2021})}\BibitemShut {NoStop}%
\bibitem [{\citenamefont {Li}\ \emph {et~al.}(2021)\citenamefont {Li}, \citenamefont {Lee},\ and\ \citenamefont {Gong}}]{Li_2021}%
  \BibitemOpen
  \bibfield  {author} {\bibinfo {author} {\bibfnamefont {L.}~\bibnamefont {Li}}, \bibinfo {author} {\bibfnamefont {C.~H.}\ \bibnamefont {Lee}}, \ and\ \bibinfo {author} {\bibfnamefont {J.}~\bibnamefont {Gong}},\ }\href {\doibase 10.1038/s42005-021-00547-x} {\bibfield  {journal} {\bibinfo  {journal} {Communications Physics}\ }\textbf {\bibinfo {volume} {4}},\ \bibinfo {pages} {42} (\bibinfo {year} {2021})}\BibitemShut {NoStop}%
\bibitem [{\citenamefont {Lin}\ \emph {et~al.}(2023)\citenamefont {Lin}, \citenamefont {Tai}, \citenamefont {Li},\ and\ \citenamefont {Lee}}]{Lin_2023}%
  \BibitemOpen
  \bibfield  {author} {\bibinfo {author} {\bibfnamefont {R.}~\bibnamefont {Lin}}, \bibinfo {author} {\bibfnamefont {T.}~\bibnamefont {Tai}}, \bibinfo {author} {\bibfnamefont {L.}~\bibnamefont {Li}}, \ and\ \bibinfo {author} {\bibfnamefont {C.~H.}\ \bibnamefont {Lee}},\ }\href {\doibase 10.1007/s11467-023-1309-z} {\bibfield  {journal} {\bibinfo  {journal} {Frontiers of Physics}\ }\textbf {\bibinfo {volume} {18}},\ \bibinfo {pages} {53605} (\bibinfo {year} {2023})}\BibitemShut {NoStop}%
\bibitem [{\citenamefont {Bergholtz}\ \emph {et~al.}(2021)\citenamefont {Bergholtz}, \citenamefont {Budich},\ and\ \citenamefont {Kunst}}]{Emil2021}%
  \BibitemOpen
  \bibfield  {author} {\bibinfo {author} {\bibfnamefont {E.~J.}\ \bibnamefont {Bergholtz}}, \bibinfo {author} {\bibfnamefont {J.~C.}\ \bibnamefont {Budich}}, \ and\ \bibinfo {author} {\bibfnamefont {F.~K.}\ \bibnamefont {Kunst}},\ }\href {\doibase 10.1103/RevModPhys.93.015005} {\bibfield  {journal} {\bibinfo  {journal} {Rev. Mod. Phys.}\ }\textbf {\bibinfo {volume} {93}},\ \bibinfo {pages} {015005} (\bibinfo {year} {2021})}\BibitemShut {NoStop}%
\bibitem [{\citenamefont {Miri}\ and\ \citenamefont {Alù}(2019)}]{Andre2019}%
  \BibitemOpen
  \bibfield  {author} {\bibinfo {author} {\bibfnamefont {M.-A.}\ \bibnamefont {Miri}}\ and\ \bibinfo {author} {\bibfnamefont {A.}~\bibnamefont {Alù}},\ }\href {\doibase 10.1126/science.aar7709} {\bibfield  {journal} {\bibinfo  {journal} {Science}\ }\textbf {\bibinfo {volume} {363}},\ \bibinfo {pages} {eaar7709} (\bibinfo {year} {2019})}\BibitemShut {NoStop}%
\bibitem [{\citenamefont {Meng}\ \emph {et~al.}(2023)\citenamefont {Meng}, \citenamefont {Ang},\ and\ \citenamefont {Lee}}]{meng2023}%
  \BibitemOpen
  \bibfield  {author} {\bibinfo {author} {\bibfnamefont {H.}~\bibnamefont {Meng}}, \bibinfo {author} {\bibfnamefont {Y.~S.}\ \bibnamefont {Ang}}, \ and\ \bibinfo {author} {\bibfnamefont {C.~H.}\ \bibnamefont {Lee}},\ }\href@noop {} {} (\bibinfo {year} {2023}),\ \Eprint {http://arxiv.org/abs/2310.16699} {arXiv:2310.16699} \BibitemShut {NoStop}%
\bibitem [{\citenamefont {{\"O}zdemir}\ \emph {et~al.}(2019)\citenamefont {{\"O}zdemir}, \citenamefont {Rotter}, \citenamefont {Nori},\ and\ \citenamefont {Yang}}]{Nori2019}%
  \BibitemOpen
  \bibfield  {author} {\bibinfo {author} {\bibfnamefont {{\c{S}}.~K.}\ \bibnamefont {{\"O}zdemir}}, \bibinfo {author} {\bibfnamefont {S.}~\bibnamefont {Rotter}}, \bibinfo {author} {\bibfnamefont {F.}~\bibnamefont {Nori}}, \ and\ \bibinfo {author} {\bibfnamefont {L.}~\bibnamefont {Yang}},\ }\href {\doibase 10.1038/s41563-019-0304-9} {\bibfield  {journal} {\bibinfo  {journal} {Nature Materials}\ }\textbf {\bibinfo {volume} {18}},\ \bibinfo {pages} {783} (\bibinfo {year} {2019})}\BibitemShut {NoStop}%
\bibitem [{\citenamefont {Krasnok}\ \emph {et~al.}(2021)\citenamefont {Krasnok}, \citenamefont {Nefedkin},\ and\ \citenamefont {Alu}}]{krasnok2021}%
  \BibitemOpen
  \bibfield  {author} {\bibinfo {author} {\bibfnamefont {A.}~\bibnamefont {Krasnok}}, \bibinfo {author} {\bibfnamefont {N.}~\bibnamefont {Nefedkin}}, \ and\ \bibinfo {author} {\bibfnamefont {A.}~\bibnamefont {Alu}},\ }\href@noop {} {} (\bibinfo {year} {2021}),\ \Eprint {http://arxiv.org/abs/2103.08135} {arXiv:2103.08135} \BibitemShut {NoStop}%
\bibitem [{\citenamefont {Kunst}\ \emph {et~al.}(2018)\citenamefont {Kunst}, \citenamefont {Edvardsson}, \citenamefont {Budich},\ and\ \citenamefont {Bergholtz}}]{Emil2018}%
  \BibitemOpen
  \bibfield  {author} {\bibinfo {author} {\bibfnamefont {F.~K.}\ \bibnamefont {Kunst}}, \bibinfo {author} {\bibfnamefont {E.}~\bibnamefont {Edvardsson}}, \bibinfo {author} {\bibfnamefont {J.~C.}\ \bibnamefont {Budich}}, \ and\ \bibinfo {author} {\bibfnamefont {E.~J.}\ \bibnamefont {Bergholtz}},\ }\href {\doibase 10.1103/PhysRevLett.121.026808} {\bibfield  {journal} {\bibinfo  {journal} {Phys. Rev. Lett.}\ }\textbf {\bibinfo {volume} {121}},\ \bibinfo {pages} {026808} (\bibinfo {year} {2018})}\BibitemShut {NoStop}%
\bibitem [{\citenamefont {Koch}\ and\ \citenamefont {Budich}(2020)}]{Koch_2020}%
  \BibitemOpen
  \bibfield  {author} {\bibinfo {author} {\bibfnamefont {R.}~\bibnamefont {Koch}}\ and\ \bibinfo {author} {\bibfnamefont {J.~C.}\ \bibnamefont {Budich}},\ }\href {\doibase 10.1140/epjd/e2020-100641-y} {\bibfield  {journal} {\bibinfo  {journal} {The European Physical Journal D}\ }\textbf {\bibinfo {volume} {74}},\ \bibinfo {pages} {70} (\bibinfo {year} {2020})}\BibitemShut {NoStop}%
\bibitem [{\citenamefont {Cao}\ \emph {et~al.}(2021)\citenamefont {Cao}, \citenamefont {Li},\ and\ \citenamefont {Yang}}]{Cao2021}%
  \BibitemOpen
  \bibfield  {author} {\bibinfo {author} {\bibfnamefont {Y.}~\bibnamefont {Cao}}, \bibinfo {author} {\bibfnamefont {Y.}~\bibnamefont {Li}}, \ and\ \bibinfo {author} {\bibfnamefont {X.}~\bibnamefont {Yang}},\ }\href {\doibase 10.1103/PhysRevB.103.075126} {\bibfield  {journal} {\bibinfo  {journal} {Phys. Rev. B}\ }\textbf {\bibinfo {volume} {103}},\ \bibinfo {pages} {075126} (\bibinfo {year} {2021})}\BibitemShut {NoStop}%
\bibitem [{\citenamefont {Xiao}\ \emph {et~al.}(2020)\citenamefont {Xiao}, \citenamefont {Deng}, \citenamefont {Wang}, \citenamefont {Zhu}, \citenamefont {Wang}, \citenamefont {Yi},\ and\ \citenamefont {Xue}}]{Xiao_2020}%
  \BibitemOpen
  \bibfield  {author} {\bibinfo {author} {\bibfnamefont {L.}~\bibnamefont {Xiao}}, \bibinfo {author} {\bibfnamefont {T.}~\bibnamefont {Deng}}, \bibinfo {author} {\bibfnamefont {K.}~\bibnamefont {Wang}}, \bibinfo {author} {\bibfnamefont {G.}~\bibnamefont {Zhu}}, \bibinfo {author} {\bibfnamefont {Z.}~\bibnamefont {Wang}}, \bibinfo {author} {\bibfnamefont {W.}~\bibnamefont {Yi}}, \ and\ \bibinfo {author} {\bibfnamefont {P.}~\bibnamefont {Xue}},\ }\href {\doibase 10.1038/s41567-020-0836-6} {\bibfield  {journal} {\bibinfo  {journal} {Nature Physics}\ }\textbf {\bibinfo {volume} {16}},\ \bibinfo {pages} {761–766} (\bibinfo {year} {2020})}\BibitemShut {NoStop}%
\bibitem [{\citenamefont {Jin}\ and\ \citenamefont {Song}(2019)}]{Jin2019}%
  \BibitemOpen
  \bibfield  {author} {\bibinfo {author} {\bibfnamefont {L.}~\bibnamefont {Jin}}\ and\ \bibinfo {author} {\bibfnamefont {Z.}~\bibnamefont {Song}},\ }\href {\doibase 10.1103/PhysRevB.99.081103} {\bibfield  {journal} {\bibinfo  {journal} {Phys. Rev. B}\ }\textbf {\bibinfo {volume} {99}},\ \bibinfo {pages} {081103} (\bibinfo {year} {2019})}\BibitemShut {NoStop}%
\bibitem [{\citenamefont {Longhi}(2019)}]{longhi_2019}%
  \BibitemOpen
  \bibfield  {author} {\bibinfo {author} {\bibfnamefont {S.}~\bibnamefont {Longhi}},\ }\href {\doibase 10.1103/PhysRevLett.122.237601} {\bibfield  {journal} {\bibinfo  {journal} {Phys. Rev. Lett.}\ }\textbf {\bibinfo {volume} {122}},\ \bibinfo {pages} {237601} (\bibinfo {year} {2019})}\BibitemShut {NoStop}%
\bibitem [{\citenamefont {Liu}\ \emph {et~al.}(2020)\citenamefont {Liu}, \citenamefont {Jiang}, \citenamefont {Cao},\ and\ \citenamefont {Chen}}]{Liu_2020}%
  \BibitemOpen
  \bibfield  {author} {\bibinfo {author} {\bibfnamefont {Y.}~\bibnamefont {Liu}}, \bibinfo {author} {\bibfnamefont {X.-P.}\ \bibnamefont {Jiang}}, \bibinfo {author} {\bibfnamefont {J.}~\bibnamefont {Cao}}, \ and\ \bibinfo {author} {\bibfnamefont {S.}~\bibnamefont {Chen}},\ }\href {\doibase 10.1103/PhysRevB.101.174205} {\bibfield  {journal} {\bibinfo  {journal} {Phys. Rev. B}\ }\textbf {\bibinfo {volume} {101}},\ \bibinfo {pages} {174205} (\bibinfo {year} {2020})}\BibitemShut {NoStop}%
\bibitem [{\citenamefont {Gandhi}\ and\ \citenamefont {Bandyopadhyay}(2023)}]{Gandhi_2023}%
  \BibitemOpen
  \bibfield  {author} {\bibinfo {author} {\bibfnamefont {S.}~\bibnamefont {Gandhi}}\ and\ \bibinfo {author} {\bibfnamefont {J.~N.}\ \bibnamefont {Bandyopadhyay}},\ }\href {\doibase 10.1103/PhysRevB.108.014204} {\bibfield  {journal} {\bibinfo  {journal} {Phys. Rev. B}\ }\textbf {\bibinfo {volume} {108}},\ \bibinfo {pages} {014204} (\bibinfo {year} {2023})}\BibitemShut {NoStop}%
\bibitem [{\citenamefont {Acharya}\ and\ \citenamefont {Datta}(2024)}]{Datta_2024}%
  \BibitemOpen
  \bibfield  {author} {\bibinfo {author} {\bibfnamefont {A.~P.}\ \bibnamefont {Acharya}}\ and\ \bibinfo {author} {\bibfnamefont {S.}~\bibnamefont {Datta}},\ }\href {\doibase 10.1103/PhysRevB.109.024203} {\bibfield  {journal} {\bibinfo  {journal} {Phys. Rev. B}\ }\textbf {\bibinfo {volume} {109}},\ \bibinfo {pages} {024203} (\bibinfo {year} {2024})}\BibitemShut {NoStop}%
\bibitem [{\citenamefont {Borgnia}\ \emph {et~al.}(2020)\citenamefont {Borgnia}, \citenamefont {Kruchkov},\ and\ \citenamefont {Slager}}]{Borgnia2020}%
  \BibitemOpen
  \bibfield  {author} {\bibinfo {author} {\bibfnamefont {D.~S.}\ \bibnamefont {Borgnia}}, \bibinfo {author} {\bibfnamefont {A.~J.}\ \bibnamefont {Kruchkov}}, \ and\ \bibinfo {author} {\bibfnamefont {R.-J.}\ \bibnamefont {Slager}},\ }\href {\doibase 10.1103/PhysRevLett.124.056802} {\bibfield  {journal} {\bibinfo  {journal} {Phys. Rev. Lett.}\ }\textbf {\bibinfo {volume} {124}},\ \bibinfo {pages} {056802} (\bibinfo {year} {2020})}\BibitemShut {NoStop}%
\bibitem [{\citenamefont {Mandal}(2024)}]{Mandal_2024}%
  \BibitemOpen
  \bibfield  {author} {\bibinfo {author} {\bibfnamefont {I.}~\bibnamefont {Mandal}},\ }\href {\doibase 10.1063/5.0198855} {\bibfield  {journal} {\bibinfo  {journal} {Journal of Applied Physics}\ }\textbf {\bibinfo {volume} {135}} (\bibinfo {year} {2024}),\ 10.1063/5.0198855}\BibitemShut {NoStop}%
\bibitem [{\citenamefont {Bender}\ and\ \citenamefont {Boettcher}(1998)}]{Bender1998}%
  \BibitemOpen
  \bibfield  {author} {\bibinfo {author} {\bibfnamefont {C.~M.}\ \bibnamefont {Bender}}\ and\ \bibinfo {author} {\bibfnamefont {S.}~\bibnamefont {Boettcher}},\ }\href {\doibase 10.1103/PhysRevLett.80.5243} {\bibfield  {journal} {\bibinfo  {journal} {Phys. Rev. Lett.}\ }\textbf {\bibinfo {volume} {80}},\ \bibinfo {pages} {5243} (\bibinfo {year} {1998})}\BibitemShut {NoStop}%
\bibitem [{\citenamefont {Bender}(2007)}]{Bender2007}%
  \BibitemOpen
  \bibfield  {author} {\bibinfo {author} {\bibfnamefont {C.~M.}\ \bibnamefont {Bender}},\ }\href {\doibase 10.1088/0034-4885/70/6/r03} {\bibfield  {journal} {\bibinfo  {journal} {Rep. Prog. Phys.}\ }\textbf {\bibinfo {volume} {70}},\ \bibinfo {pages} {947} (\bibinfo {year} {2007})}\BibitemShut {NoStop}%
\bibitem [{\citenamefont {Hatano}\ and\ \citenamefont {Nelson}(1996)}]{Nelson_1996}%
  \BibitemOpen
  \bibfield  {author} {\bibinfo {author} {\bibfnamefont {N.}~\bibnamefont {Hatano}}\ and\ \bibinfo {author} {\bibfnamefont {D.~R.}\ \bibnamefont {Nelson}},\ }\href {\doibase 10.1103/PhysRevLett.77.570} {\bibfield  {journal} {\bibinfo  {journal} {Phys. Rev. Lett.}\ }\textbf {\bibinfo {volume} {77}},\ \bibinfo {pages} {570} (\bibinfo {year} {1996})}\BibitemShut {NoStop}%
\bibitem [{\citenamefont {Hatano}\ and\ \citenamefont {Nelson}(1997)}]{Nelson_1997}%
  \BibitemOpen
  \bibfield  {author} {\bibinfo {author} {\bibfnamefont {N.}~\bibnamefont {Hatano}}\ and\ \bibinfo {author} {\bibfnamefont {D.~R.}\ \bibnamefont {Nelson}},\ }\href {\doibase 10.1103/PhysRevB.56.8651} {\bibfield  {journal} {\bibinfo  {journal} {Phys. Rev. B}\ }\textbf {\bibinfo {volume} {56}},\ \bibinfo {pages} {8651} (\bibinfo {year} {1997})}\BibitemShut {NoStop}%
\bibitem [{\citenamefont {Hatano}\ and\ \citenamefont {Nelson}(1998)}]{Nelson1998}%
  \BibitemOpen
  \bibfield  {author} {\bibinfo {author} {\bibfnamefont {N.}~\bibnamefont {Hatano}}\ and\ \bibinfo {author} {\bibfnamefont {D.~R.}\ \bibnamefont {Nelson}},\ }\href {\doibase 10.1103/PhysRevB.58.8384} {\bibfield  {journal} {\bibinfo  {journal} {Phys. Rev. B}\ }\textbf {\bibinfo {volume} {58}},\ \bibinfo {pages} {8384} (\bibinfo {year} {1998})}\BibitemShut {NoStop}%
\bibitem [{\citenamefont {Hamazaki}\ \emph {et~al.}(2019)\citenamefont {Hamazaki}, \citenamefont {Kawabata},\ and\ \citenamefont {Ueda}}]{Ueda2019}%
  \BibitemOpen
  \bibfield  {author} {\bibinfo {author} {\bibfnamefont {R.}~\bibnamefont {Hamazaki}}, \bibinfo {author} {\bibfnamefont {K.}~\bibnamefont {Kawabata}}, \ and\ \bibinfo {author} {\bibfnamefont {M.}~\bibnamefont {Ueda}},\ }\href {\doibase 10.1103/PhysRevLett.123.090603} {\bibfield  {journal} {\bibinfo  {journal} {Phys. Rev. Lett.}\ }\textbf {\bibinfo {volume} {123}},\ \bibinfo {pages} {090603} (\bibinfo {year} {2019})}\BibitemShut {NoStop}%
\bibitem [{\citenamefont {Zhang}\ \emph {et~al.}(2022)\citenamefont {Zhang}, \citenamefont {Denner}, \citenamefont {Bzdu\ifmmode~\check{s}\else \v{s}\fi{}ek}, \citenamefont {Sentef},\ and\ \citenamefont {Neupert}}]{Titus2022}%
  \BibitemOpen
  \bibfield  {author} {\bibinfo {author} {\bibfnamefont {S.-B.}\ \bibnamefont {Zhang}}, \bibinfo {author} {\bibfnamefont {M.~M.}\ \bibnamefont {Denner}}, \bibinfo {author} {\bibfnamefont {T.~c.~v.}\ \bibnamefont {Bzdu\ifmmode~\check{s}\else \v{s}\fi{}ek}}, \bibinfo {author} {\bibfnamefont {M.~A.}\ \bibnamefont {Sentef}}, \ and\ \bibinfo {author} {\bibfnamefont {T.}~\bibnamefont {Neupert}},\ }\href {\doibase 10.1103/PhysRevB.106.L121102} {\bibfield  {journal} {\bibinfo  {journal} {Phys. Rev. B}\ }\textbf {\bibinfo {volume} {106}},\ \bibinfo {pages} {L121102} (\bibinfo {year} {2022})}\BibitemShut {NoStop}%
\bibitem [{\citenamefont {Zhai}\ \emph {et~al.}(2020)\citenamefont {Zhai}, \citenamefont {Yin},\ and\ \citenamefont {Huang}}]{Huang2020}%
  \BibitemOpen
  \bibfield  {author} {\bibinfo {author} {\bibfnamefont {L.-J.}\ \bibnamefont {Zhai}}, \bibinfo {author} {\bibfnamefont {S.}~\bibnamefont {Yin}}, \ and\ \bibinfo {author} {\bibfnamefont {G.-Y.}\ \bibnamefont {Huang}},\ }\href {\doibase 10.1103/PhysRevB.102.064206} {\bibfield  {journal} {\bibinfo  {journal} {Phys. Rev. B}\ }\textbf {\bibinfo {volume} {102}},\ \bibinfo {pages} {064206} (\bibinfo {year} {2020})}\BibitemShut {NoStop}%
\bibitem [{\citenamefont {Mu}\ \emph {et~al.}(2020)\citenamefont {Mu}, \citenamefont {Lee}, \citenamefont {Li},\ and\ \citenamefont {Gong}}]{Gong2020}%
  \BibitemOpen
  \bibfield  {author} {\bibinfo {author} {\bibfnamefont {S.}~\bibnamefont {Mu}}, \bibinfo {author} {\bibfnamefont {C.~H.}\ \bibnamefont {Lee}}, \bibinfo {author} {\bibfnamefont {L.}~\bibnamefont {Li}}, \ and\ \bibinfo {author} {\bibfnamefont {J.}~\bibnamefont {Gong}},\ }\href {\doibase 10.1103/PhysRevB.102.081115} {\bibfield  {journal} {\bibinfo  {journal} {Phys. Rev. B}\ }\textbf {\bibinfo {volume} {102}},\ \bibinfo {pages} {081115} (\bibinfo {year} {2020})}\BibitemShut {NoStop}%
\bibitem [{\citenamefont {Heu\ss{}en}\ \emph {et~al.}(2021)\citenamefont {Heu\ss{}en}, \citenamefont {White},\ and\ \citenamefont {Refael}}]{Gil2021}%
  \BibitemOpen
  \bibfield  {author} {\bibinfo {author} {\bibfnamefont {S.}~\bibnamefont {Heu\ss{}en}}, \bibinfo {author} {\bibfnamefont {C.~D.}\ \bibnamefont {White}}, \ and\ \bibinfo {author} {\bibfnamefont {G.}~\bibnamefont {Refael}},\ }\href {\doibase 10.1103/PhysRevB.103.064201} {\bibfield  {journal} {\bibinfo  {journal} {Phys. Rev. B}\ }\textbf {\bibinfo {volume} {103}},\ \bibinfo {pages} {064201} (\bibinfo {year} {2021})}\BibitemShut {NoStop}%
\bibitem [{\citenamefont {Orito}\ and\ \citenamefont {Imura}(2022)}]{Ichiro_2022}%
  \BibitemOpen
  \bibfield  {author} {\bibinfo {author} {\bibfnamefont {T.}~\bibnamefont {Orito}}\ and\ \bibinfo {author} {\bibfnamefont {K.-I.}\ \bibnamefont {Imura}},\ }\href {\doibase 10.1103/PhysRevB.105.024303} {\bibfield  {journal} {\bibinfo  {journal} {Phys. Rev. B}\ }\textbf {\bibinfo {volume} {105}},\ \bibinfo {pages} {024303} (\bibinfo {year} {2022})}\BibitemShut {NoStop}%
\bibitem [{\citenamefont {Zeng}(2022{\natexlab{b}})}]{QiBoZeng2022}%
  \BibitemOpen
  \bibfield  {author} {\bibinfo {author} {\bibfnamefont {Q.-B.}\ \bibnamefont {Zeng}},\ }\href {\doibase 10.1103/PhysRevB.106.235411} {\bibfield  {journal} {\bibinfo  {journal} {Phys. Rev. B}\ }\textbf {\bibinfo {volume} {106}},\ \bibinfo {pages} {235411} (\bibinfo {year} {2022}{\natexlab{b}})}\BibitemShut {NoStop}%
\bibitem [{\citenamefont {Liu}\ \emph {et~al.}(2021{\natexlab{a}})\citenamefont {Liu}, \citenamefont {Shao}, \citenamefont {Ma}, \citenamefont {Zhang}, \citenamefont {You}, \citenamefont {Wu}, \citenamefont {Xiang}, \citenamefont {Cui},\ and\ \citenamefont {Zhang}}]{Liu2021}%
  \BibitemOpen
  \bibfield  {author} {\bibinfo {author} {\bibfnamefont {S.}~\bibnamefont {Liu}}, \bibinfo {author} {\bibfnamefont {R.}~\bibnamefont {Shao}}, \bibinfo {author} {\bibfnamefont {S.}~\bibnamefont {Ma}}, \bibinfo {author} {\bibfnamefont {L.}~\bibnamefont {Zhang}}, \bibinfo {author} {\bibfnamefont {O.}~\bibnamefont {You}}, \bibinfo {author} {\bibfnamefont {H.}~\bibnamefont {Wu}}, \bibinfo {author} {\bibfnamefont {Y.~J.}\ \bibnamefont {Xiang}}, \bibinfo {author} {\bibfnamefont {T.~J.}\ \bibnamefont {Cui}}, \ and\ \bibinfo {author} {\bibfnamefont {S.}~\bibnamefont {Zhang}},\ }\href {\doibase 10.34133/2021/5608038} {\bibfield  {journal} {\bibinfo  {journal} {Research}\ }\textbf {\bibinfo {volume} {2021}} (\bibinfo {year} {2021}{\natexlab{a}}),\ 10.34133/2021/5608038}\BibitemShut {NoStop}%
\bibitem [{\citenamefont {Zhang}\ \emph {et~al.}(2023)\citenamefont {Zhang}, \citenamefont {Chen}, \citenamefont {Li}, \citenamefont {Lee},\ and\ \citenamefont {Zhang}}]{Zhang2023}%
  \BibitemOpen
  \bibfield  {author} {\bibinfo {author} {\bibfnamefont {H.}~\bibnamefont {Zhang}}, \bibinfo {author} {\bibfnamefont {T.}~\bibnamefont {Chen}}, \bibinfo {author} {\bibfnamefont {L.}~\bibnamefont {Li}}, \bibinfo {author} {\bibfnamefont {C.~H.}\ \bibnamefont {Lee}}, \ and\ \bibinfo {author} {\bibfnamefont {X.}~\bibnamefont {Zhang}},\ }\href {\doibase 10.1103/PhysRevB.107.085426} {\bibfield  {journal} {\bibinfo  {journal} {Phys. Rev. B}\ }\textbf {\bibinfo {volume} {107}},\ \bibinfo {pages} {085426} (\bibinfo {year} {2023})}\BibitemShut {NoStop}%
\bibitem [{\citenamefont {Guo}\ \emph {et~al.}(2022)\citenamefont {Guo}, \citenamefont {Dong}, \citenamefont {Zhang}, \citenamefont {Hu},\ and\ \citenamefont {Yang}}]{Dong}%
  \BibitemOpen
  \bibfield  {author} {\bibinfo {author} {\bibfnamefont {S.}~\bibnamefont {Guo}}, \bibinfo {author} {\bibfnamefont {C.}~\bibnamefont {Dong}}, \bibinfo {author} {\bibfnamefont {F.}~\bibnamefont {Zhang}}, \bibinfo {author} {\bibfnamefont {J.}~\bibnamefont {Hu}}, \ and\ \bibinfo {author} {\bibfnamefont {Z.}~\bibnamefont {Yang}},\ }\href {\doibase 10.1103/PhysRevA.106.L061302} {\bibfield  {journal} {\bibinfo  {journal} {Phys. Rev. A}\ }\textbf {\bibinfo {volume} {106}},\ \bibinfo {pages} {L061302} (\bibinfo {year} {2022})}\BibitemShut {NoStop}%
\bibitem [{\citenamefont {{Zhou}}\ \emph {et~al.}(2022)\citenamefont {{Zhou}}, \citenamefont {{Li}}, \citenamefont {{Yi}},\ and\ \citenamefont {{Cui}}}]{Zhou}%
  \BibitemOpen
  \bibfield  {author} {\bibinfo {author} {\bibfnamefont {L.}~\bibnamefont {{Zhou}}}, \bibinfo {author} {\bibfnamefont {H.}~\bibnamefont {{Li}}}, \bibinfo {author} {\bibfnamefont {W.}~\bibnamefont {{Yi}}}, \ and\ \bibinfo {author} {\bibfnamefont {X.}~\bibnamefont {{Cui}}},\ }\href {\doibase 10.1038/s42005-022-01021-y} {\bibfield  {journal} {\bibinfo  {journal} {Communications Physics}\ }\textbf {\bibinfo {volume} {5}},\ \bibinfo {eid} {252} (\bibinfo {year} {2022})}\BibitemShut {NoStop}%
\bibitem [{\citenamefont {Li}\ \emph {et~al.}(2020{\natexlab{b}})\citenamefont {Li}, \citenamefont {Lee},\ and\ \citenamefont {Gong}}]{Gong}%
  \BibitemOpen
  \bibfield  {author} {\bibinfo {author} {\bibfnamefont {L.}~\bibnamefont {Li}}, \bibinfo {author} {\bibfnamefont {C.~H.}\ \bibnamefont {Lee}}, \ and\ \bibinfo {author} {\bibfnamefont {J.}~\bibnamefont {Gong}},\ }\href {\doibase 10.1103/PhysRevLett.124.250402} {\bibfield  {journal} {\bibinfo  {journal} {Phys. Rev. Lett.}\ }\textbf {\bibinfo {volume} {124}},\ \bibinfo {pages} {250402} (\bibinfo {year} {2020}{\natexlab{b}})}\BibitemShut {NoStop}%
\bibitem [{\citenamefont {Liang}\ \emph {et~al.}(2022)\citenamefont {Liang}, \citenamefont {Xie}, \citenamefont {Dong}, \citenamefont {Li}, \citenamefont {Li}, \citenamefont {Gadway}, \citenamefont {Yi},\ and\ \citenamefont {Yan}}]{BoYan}%
  \BibitemOpen
  \bibfield  {author} {\bibinfo {author} {\bibfnamefont {Q.}~\bibnamefont {Liang}}, \bibinfo {author} {\bibfnamefont {D.}~\bibnamefont {Xie}}, \bibinfo {author} {\bibfnamefont {Z.}~\bibnamefont {Dong}}, \bibinfo {author} {\bibfnamefont {H.}~\bibnamefont {Li}}, \bibinfo {author} {\bibfnamefont {H.}~\bibnamefont {Li}}, \bibinfo {author} {\bibfnamefont {B.}~\bibnamefont {Gadway}}, \bibinfo {author} {\bibfnamefont {W.}~\bibnamefont {Yi}}, \ and\ \bibinfo {author} {\bibfnamefont {B.}~\bibnamefont {Yan}},\ }\href {\doibase 10.1103/PhysRevLett.129.070401} {\bibfield  {journal} {\bibinfo  {journal} {Phys. Rev. Lett.}\ }\textbf {\bibinfo {volume} {129}},\ \bibinfo {pages} {070401} (\bibinfo {year} {2022})}\BibitemShut {NoStop}%
\bibitem [{\citenamefont {{Zhang}}\ \emph {et~al.}(2021)\citenamefont {{Zhang}}, \citenamefont {{Yang}}, \citenamefont {{Ge}}, \citenamefont {{Guan}}, \citenamefont {{Chen}}, \citenamefont {{Yan}}, \citenamefont {{Chen}}, \citenamefont {{Xi}}, \citenamefont {{Li}}, \citenamefont {{Jia}}, \citenamefont {{Yuan}}, \citenamefont {{Sun}}, \citenamefont {{Chen}},\ and\ \citenamefont {{Zhang}}}]{Zhang}%
  \BibitemOpen
  \bibfield  {author} {\bibinfo {author} {\bibfnamefont {L.}~\bibnamefont {{Zhang}}}, \bibinfo {author} {\bibfnamefont {Y.}~\bibnamefont {{Yang}}}, \bibinfo {author} {\bibfnamefont {Y.}~\bibnamefont {{Ge}}}, \bibinfo {author} {\bibfnamefont {Y.-J.}\ \bibnamefont {{Guan}}}, \bibinfo {author} {\bibfnamefont {Q.}~\bibnamefont {{Chen}}}, \bibinfo {author} {\bibfnamefont {Q.}~\bibnamefont {{Yan}}}, \bibinfo {author} {\bibfnamefont {F.}~\bibnamefont {{Chen}}}, \bibinfo {author} {\bibfnamefont {R.}~\bibnamefont {{Xi}}}, \bibinfo {author} {\bibfnamefont {Y.}~\bibnamefont {{Li}}}, \bibinfo {author} {\bibfnamefont {D.}~\bibnamefont {{Jia}}}, \bibinfo {author} {\bibfnamefont {S.-Q.}\ \bibnamefont {{Yuan}}}, \bibinfo {author} {\bibfnamefont {H.-X.}\ \bibnamefont {{Sun}}}, \bibinfo {author} {\bibfnamefont {H.}~\bibnamefont {{Chen}}}, \ and\ \bibinfo {author} {\bibfnamefont {B.}~\bibnamefont {{Zhang}}},\ }\href {\doibase 10.1038/s41467-021-26619-8} {\bibfield  {journal} {\bibinfo  {journal} {Nature Communications}\
  }\textbf {\bibinfo {volume} {12}},\ \bibinfo {eid} {6297} (\bibinfo {year} {2021})}\BibitemShut {NoStop}%
\bibitem [{\citenamefont {Wang}\ \emph {et~al.}(2021{\natexlab{a}})\citenamefont {Wang}, \citenamefont {Dutt}, \citenamefont {Yang}, \citenamefont {Wojcik}, \citenamefont {Vučković},\ and\ \citenamefont {Fan}}]{Wang_2021}%
  \BibitemOpen
  \bibfield  {author} {\bibinfo {author} {\bibfnamefont {K.}~\bibnamefont {Wang}}, \bibinfo {author} {\bibfnamefont {A.}~\bibnamefont {Dutt}}, \bibinfo {author} {\bibfnamefont {K.~Y.}\ \bibnamefont {Yang}}, \bibinfo {author} {\bibfnamefont {C.~C.}\ \bibnamefont {Wojcik}}, \bibinfo {author} {\bibfnamefont {J.}~\bibnamefont {Vučković}}, \ and\ \bibinfo {author} {\bibfnamefont {S.}~\bibnamefont {Fan}},\ }\href {\doibase 10.1126/science.abf6568} {\bibfield  {journal} {\bibinfo  {journal} {Science}\ }\textbf {\bibinfo {volume} {371}},\ \bibinfo {pages} {1240–1245} (\bibinfo {year} {2021}{\natexlab{a}})}\BibitemShut {NoStop}%
\bibitem [{\citenamefont {Lin}\ \emph {et~al.}(2022)\citenamefont {Lin}, \citenamefont {Li}, \citenamefont {Xiao}, \citenamefont {Wang}, \citenamefont {Yi},\ and\ \citenamefont {Xue}}]{Quan}%
  \BibitemOpen
  \bibfield  {author} {\bibinfo {author} {\bibfnamefont {Q.}~\bibnamefont {Lin}}, \bibinfo {author} {\bibfnamefont {T.}~\bibnamefont {Li}}, \bibinfo {author} {\bibfnamefont {L.}~\bibnamefont {Xiao}}, \bibinfo {author} {\bibfnamefont {K.}~\bibnamefont {Wang}}, \bibinfo {author} {\bibfnamefont {W.}~\bibnamefont {Yi}}, \ and\ \bibinfo {author} {\bibfnamefont {P.}~\bibnamefont {Xue}},\ }\href {\doibase 10.1103/PhysRevLett.129.113601} {\bibfield  {journal} {\bibinfo  {journal} {Phys. Rev. Lett.}\ }\textbf {\bibinfo {volume} {129}},\ \bibinfo {pages} {113601} (\bibinfo {year} {2022})}\BibitemShut {NoStop}%
\bibitem [{\citenamefont {Wang}\ \emph {et~al.}(2021{\natexlab{b}})\citenamefont {Wang}, \citenamefont {Li}, \citenamefont {Xiao}, \citenamefont {Han}, \citenamefont {Yi},\ and\ \citenamefont {Xue}}]{PhysRevLettKunkun}%
  \BibitemOpen
  \bibfield  {author} {\bibinfo {author} {\bibfnamefont {K.}~\bibnamefont {Wang}}, \bibinfo {author} {\bibfnamefont {T.}~\bibnamefont {Li}}, \bibinfo {author} {\bibfnamefont {L.}~\bibnamefont {Xiao}}, \bibinfo {author} {\bibfnamefont {Y.}~\bibnamefont {Han}}, \bibinfo {author} {\bibfnamefont {W.}~\bibnamefont {Yi}}, \ and\ \bibinfo {author} {\bibfnamefont {P.}~\bibnamefont {Xue}},\ }\href {\doibase 10.1103/PhysRevLett.127.270602} {\bibfield  {journal} {\bibinfo  {journal} {Phys. Rev. Lett.}\ }\textbf {\bibinfo {volume} {127}},\ \bibinfo {pages} {270602} (\bibinfo {year} {2021}{\natexlab{b}})}\BibitemShut {NoStop}%
\bibitem [{\citenamefont {Ghatak}\ \emph {et~al.}(2020)\citenamefont {Ghatak}, \citenamefont {Brandenbourger}, \citenamefont {van Wezel},\ and\ \citenamefont {Coulais}}]{mechanicalmetamaterial}%
  \BibitemOpen
  \bibfield  {author} {\bibinfo {author} {\bibfnamefont {A.}~\bibnamefont {Ghatak}}, \bibinfo {author} {\bibfnamefont {M.}~\bibnamefont {Brandenbourger}}, \bibinfo {author} {\bibfnamefont {J.}~\bibnamefont {van Wezel}}, \ and\ \bibinfo {author} {\bibfnamefont {C.}~\bibnamefont {Coulais}},\ }\href {\doibase 10.1073/pnas.2010580117} {\bibfield  {journal} {\bibinfo  {journal} {Proceedings of the National Academy of Sciences}\ }\textbf {\bibinfo {volume} {117}},\ \bibinfo {pages} {29561} (\bibinfo {year} {2020})}\BibitemShut {NoStop}%
\bibitem [{\citenamefont {Longhi}(2021{\natexlab{a}})}]{longhi2021}%
  \BibitemOpen
  \bibfield  {author} {\bibinfo {author} {\bibfnamefont {S.}~\bibnamefont {Longhi}},\ }\href {\doibase 10.1103/PhysRevB.103.054203} {\bibfield  {journal} {\bibinfo  {journal} {Phys. Rev. B}\ }\textbf {\bibinfo {volume} {103}},\ \bibinfo {pages} {054203} (\bibinfo {year} {2021}{\natexlab{a}})}\BibitemShut {NoStop}%
\bibitem [{\citenamefont {Tang}\ \emph {et~al.}(2021)\citenamefont {Tang}, \citenamefont {Zhang}, \citenamefont {Zhang},\ and\ \citenamefont {Zhang}}]{Wei2021}%
  \BibitemOpen
  \bibfield  {author} {\bibinfo {author} {\bibfnamefont {L.-Z.}\ \bibnamefont {Tang}}, \bibinfo {author} {\bibfnamefont {G.-Q.}\ \bibnamefont {Zhang}}, \bibinfo {author} {\bibfnamefont {L.-F.}\ \bibnamefont {Zhang}}, \ and\ \bibinfo {author} {\bibfnamefont {D.-W.}\ \bibnamefont {Zhang}},\ }\href {\doibase 10.1103/PhysRevA.103.033325} {\bibfield  {journal} {\bibinfo  {journal} {Phys. Rev. A}\ }\textbf {\bibinfo {volume} {103}},\ \bibinfo {pages} {033325} (\bibinfo {year} {2021})}\BibitemShut {NoStop}%
\bibitem [{\citenamefont {Cai}(2022)}]{cai2022}%
  \BibitemOpen
  \bibfield  {author} {\bibinfo {author} {\bibfnamefont {X.}~\bibnamefont {Cai}},\ }\href {\doibase 10.1103/PhysRevB.106.214207} {\bibfield  {journal} {\bibinfo  {journal} {Phys. Rev. B}\ }\textbf {\bibinfo {volume} {106}},\ \bibinfo {pages} {214207} (\bibinfo {year} {2022})}\BibitemShut {NoStop}%
\bibitem [{\citenamefont {Liu}\ \emph {et~al.}(2021{\natexlab{b}})\citenamefont {Liu}, \citenamefont {Zhou},\ and\ \citenamefont {Chen}}]{shu2021}%
  \BibitemOpen
  \bibfield  {author} {\bibinfo {author} {\bibfnamefont {Y.}~\bibnamefont {Liu}}, \bibinfo {author} {\bibfnamefont {Q.}~\bibnamefont {Zhou}}, \ and\ \bibinfo {author} {\bibfnamefont {S.}~\bibnamefont {Chen}},\ }\href {\doibase 10.1103/PhysRevB.104.024201} {\bibfield  {journal} {\bibinfo  {journal} {Phys. Rev. B}\ }\textbf {\bibinfo {volume} {104}},\ \bibinfo {pages} {024201} (\bibinfo {year} {2021}{\natexlab{b}})}\BibitemShut {NoStop}%
\bibitem [{\citenamefont {Zeng}\ \emph {et~al.}(2020{\natexlab{a}})\citenamefont {Zeng}, \citenamefont {Yang},\ and\ \citenamefont {Xu}}]{Zeng2020}%
  \BibitemOpen
  \bibfield  {author} {\bibinfo {author} {\bibfnamefont {Q.-B.}\ \bibnamefont {Zeng}}, \bibinfo {author} {\bibfnamefont {Y.-B.}\ \bibnamefont {Yang}}, \ and\ \bibinfo {author} {\bibfnamefont {Y.}~\bibnamefont {Xu}},\ }\href {\doibase 10.1103/PhysRevB.101.020201} {\bibfield  {journal} {\bibinfo  {journal} {Phys. Rev. B}\ }\textbf {\bibinfo {volume} {101}},\ \bibinfo {pages} {020201} (\bibinfo {year} {2020}{\natexlab{a}})}\BibitemShut {NoStop}%
\bibitem [{\citenamefont {Zhou}(2023)}]{Zhou2023}%
  \BibitemOpen
  \bibfield  {author} {\bibinfo {author} {\bibfnamefont {L.}~\bibnamefont {Zhou}},\ }\href {\doibase 10.1103/PhysRevB.108.014202} {\bibfield  {journal} {\bibinfo  {journal} {Phys. Rev. B}\ }\textbf {\bibinfo {volume} {108}},\ \bibinfo {pages} {014202} (\bibinfo {year} {2023})}\BibitemShut {NoStop}%
\bibitem [{\citenamefont {Wang}\ \emph {et~al.}(2023)\citenamefont {Wang}, \citenamefont {Zheng}, \citenamefont {Chen}, \citenamefont {Xiao}, \citenamefont {Jia},\ and\ \citenamefont {Zhang}}]{Jia2023}%
  \BibitemOpen
  \bibfield  {author} {\bibinfo {author} {\bibfnamefont {H.}~\bibnamefont {Wang}}, \bibinfo {author} {\bibfnamefont {X.}~\bibnamefont {Zheng}}, \bibinfo {author} {\bibfnamefont {J.}~\bibnamefont {Chen}}, \bibinfo {author} {\bibfnamefont {L.}~\bibnamefont {Xiao}}, \bibinfo {author} {\bibfnamefont {S.}~\bibnamefont {Jia}}, \ and\ \bibinfo {author} {\bibfnamefont {L.}~\bibnamefont {Zhang}},\ }\href {\doibase 10.1103/PhysRevB.107.075128} {\bibfield  {journal} {\bibinfo  {journal} {Phys. Rev. B}\ }\textbf {\bibinfo {volume} {107}},\ \bibinfo {pages} {075128} (\bibinfo {year} {2023})}\BibitemShut {NoStop}%
\bibitem [{\citenamefont {Yuce}\ and\ \citenamefont {Ramezani}(2022)}]{Yuce2022}%
  \BibitemOpen
  \bibfield  {author} {\bibinfo {author} {\bibfnamefont {C.}~\bibnamefont {Yuce}}\ and\ \bibinfo {author} {\bibfnamefont {H.}~\bibnamefont {Ramezani}},\ }\href {\doibase 10.1103/PhysRevB.106.024202} {\bibfield  {journal} {\bibinfo  {journal} {Phys. Rev. B}\ }\textbf {\bibinfo {volume} {106}},\ \bibinfo {pages} {024202} (\bibinfo {year} {2022})}\BibitemShut {NoStop}%
\bibitem [{\citenamefont {Chen}\ \emph {et~al.}(2022)\citenamefont {Chen}, \citenamefont {Zhou}, \citenamefont {Chen},\ and\ \citenamefont {Ye}}]{Peng2022}%
  \BibitemOpen
  \bibfield  {author} {\bibinfo {author} {\bibfnamefont {L.-M.}\ \bibnamefont {Chen}}, \bibinfo {author} {\bibfnamefont {Y.}~\bibnamefont {Zhou}}, \bibinfo {author} {\bibfnamefont {S.~A.}\ \bibnamefont {Chen}}, \ and\ \bibinfo {author} {\bibfnamefont {P.}~\bibnamefont {Ye}},\ }\href {\doibase 10.1103/PhysRevB.105.L121115} {\bibfield  {journal} {\bibinfo  {journal} {Phys. Rev. B}\ }\textbf {\bibinfo {volume} {105}},\ \bibinfo {pages} {L121115} (\bibinfo {year} {2022})}\BibitemShut {NoStop}%
\bibitem [{\citenamefont {Claes}\ and\ \citenamefont {Hughes}(2021)}]{Taylor2021}%
  \BibitemOpen
  \bibfield  {author} {\bibinfo {author} {\bibfnamefont {J.}~\bibnamefont {Claes}}\ and\ \bibinfo {author} {\bibfnamefont {T.~L.}\ \bibnamefont {Hughes}},\ }\href {\doibase 10.1103/PhysRevB.103.L140201} {\bibfield  {journal} {\bibinfo  {journal} {Phys. Rev. B}\ }\textbf {\bibinfo {volume} {103}},\ \bibinfo {pages} {L140201} (\bibinfo {year} {2021})}\BibitemShut {NoStop}%
\bibitem [{\citenamefont {{Wu}}\ \emph {et~al.}(2021)\citenamefont {{Wu}}, \citenamefont {{Fan}}, \citenamefont {{Chen}},\ and\ \citenamefont {{Jia}}}]{Jia2021}%
  \BibitemOpen
  \bibfield  {author} {\bibinfo {author} {\bibfnamefont {C.}~\bibnamefont {{Wu}}}, \bibinfo {author} {\bibfnamefont {J.}~\bibnamefont {{Fan}}}, \bibinfo {author} {\bibfnamefont {G.}~\bibnamefont {{Chen}}}, \ and\ \bibinfo {author} {\bibfnamefont {S.}~\bibnamefont {{Jia}}},\ }\href {\doibase 10.1088/1367-2630/ac430b} {\bibfield  {journal} {\bibinfo  {journal} {New Journal of Physics}\ }\textbf {\bibinfo {volume} {23}},\ \bibinfo {eid} {123048} (\bibinfo {year} {2021})}\BibitemShut {NoStop}%
\bibitem [{\citenamefont {Jiang}\ \emph {et~al.}(2019)\citenamefont {Jiang}, \citenamefont {Lang}, \citenamefont {Yang}, \citenamefont {Zhu},\ and\ \citenamefont {Chen}}]{shuchen_2019}%
  \BibitemOpen
  \bibfield  {author} {\bibinfo {author} {\bibfnamefont {H.}~\bibnamefont {Jiang}}, \bibinfo {author} {\bibfnamefont {L.-J.}\ \bibnamefont {Lang}}, \bibinfo {author} {\bibfnamefont {C.}~\bibnamefont {Yang}}, \bibinfo {author} {\bibfnamefont {S.-L.}\ \bibnamefont {Zhu}}, \ and\ \bibinfo {author} {\bibfnamefont {S.}~\bibnamefont {Chen}},\ }\href {\doibase 10.1103/PhysRevB.100.054301} {\bibfield  {journal} {\bibinfo  {journal} {Phys. Rev. B}\ }\textbf {\bibinfo {volume} {100}},\ \bibinfo {pages} {054301} (\bibinfo {year} {2019})}\BibitemShut {NoStop}%
\bibitem [{\citenamefont {Zeng}\ \emph {et~al.}(2020{\natexlab{b}})\citenamefont {Zeng}, \citenamefont {Yang},\ and\ \citenamefont {L\"u}}]{Rong_2020}%
  \BibitemOpen
  \bibfield  {author} {\bibinfo {author} {\bibfnamefont {Q.-B.}\ \bibnamefont {Zeng}}, \bibinfo {author} {\bibfnamefont {Y.-B.}\ \bibnamefont {Yang}}, \ and\ \bibinfo {author} {\bibfnamefont {R.}~\bibnamefont {L\"u}},\ }\href {\doibase 10.1103/PhysRevB.101.125418} {\bibfield  {journal} {\bibinfo  {journal} {Phys. Rev. B}\ }\textbf {\bibinfo {volume} {101}},\ \bibinfo {pages} {125418} (\bibinfo {year} {2020}{\natexlab{b}})}\BibitemShut {NoStop}%
\bibitem [{\citenamefont {Liu}\ \emph {et~al.}(2021{\natexlab{c}})\citenamefont {Liu}, \citenamefont {Wang}, \citenamefont {Liu}, \citenamefont {Zhou},\ and\ \citenamefont {Chen}}]{Liu1_2021}%
  \BibitemOpen
  \bibfield  {author} {\bibinfo {author} {\bibfnamefont {Y.}~\bibnamefont {Liu}}, \bibinfo {author} {\bibfnamefont {Y.}~\bibnamefont {Wang}}, \bibinfo {author} {\bibfnamefont {X.-J.}\ \bibnamefont {Liu}}, \bibinfo {author} {\bibfnamefont {Q.}~\bibnamefont {Zhou}}, \ and\ \bibinfo {author} {\bibfnamefont {S.}~\bibnamefont {Chen}},\ }\href {\doibase 10.1103/PhysRevB.103.014203} {\bibfield  {journal} {\bibinfo  {journal} {Phys. Rev. B}\ }\textbf {\bibinfo {volume} {103}},\ \bibinfo {pages} {014203} (\bibinfo {year} {2021}{\natexlab{c}})}\BibitemShut {NoStop}%
\bibitem [{\citenamefont {Liu}\ \emph {et~al.}(2021{\natexlab{d}})\citenamefont {Liu}, \citenamefont {Wang}, \citenamefont {Zheng},\ and\ \citenamefont {Chen}}]{Liu2_2021}%
  \BibitemOpen
  \bibfield  {author} {\bibinfo {author} {\bibfnamefont {Y.}~\bibnamefont {Liu}}, \bibinfo {author} {\bibfnamefont {Y.}~\bibnamefont {Wang}}, \bibinfo {author} {\bibfnamefont {Z.}~\bibnamefont {Zheng}}, \ and\ \bibinfo {author} {\bibfnamefont {S.}~\bibnamefont {Chen}},\ }\href {\doibase 10.1103/PhysRevB.103.134208} {\bibfield  {journal} {\bibinfo  {journal} {Phys. Rev. B}\ }\textbf {\bibinfo {volume} {103}},\ \bibinfo {pages} {134208} (\bibinfo {year} {2021}{\natexlab{d}})}\BibitemShut {NoStop}%
\bibitem [{\citenamefont {Cai}(2021)}]{Cai_2021}%
  \BibitemOpen
  \bibfield  {author} {\bibinfo {author} {\bibfnamefont {X.}~\bibnamefont {Cai}},\ }\href {\doibase 10.1103/PhysRevB.103.014201} {\bibfield  {journal} {\bibinfo  {journal} {Phys. Rev. B}\ }\textbf {\bibinfo {volume} {103}},\ \bibinfo {pages} {014201} (\bibinfo {year} {2021})}\BibitemShut {NoStop}%
\bibitem [{\citenamefont {Liu}\ \emph {et~al.}(2021{\natexlab{e}})\citenamefont {Liu}, \citenamefont {Cheng}, \citenamefont {Guo},\ and\ \citenamefont {Xianlong}}]{Gao_2021}%
  \BibitemOpen
  \bibfield  {author} {\bibinfo {author} {\bibfnamefont {T.}~\bibnamefont {Liu}}, \bibinfo {author} {\bibfnamefont {S.}~\bibnamefont {Cheng}}, \bibinfo {author} {\bibfnamefont {H.}~\bibnamefont {Guo}}, \ and\ \bibinfo {author} {\bibfnamefont {G.}~\bibnamefont {Xianlong}},\ }\href {\doibase 10.1103/PhysRevB.103.104203} {\bibfield  {journal} {\bibinfo  {journal} {Phys. Rev. B}\ }\textbf {\bibinfo {volume} {103}},\ \bibinfo {pages} {104203} (\bibinfo {year} {2021}{\natexlab{e}})}\BibitemShut {NoStop}%
\bibitem [{\citenamefont {Wang}\ \emph {et~al.}(2021{\natexlab{c}})\citenamefont {Wang}, \citenamefont {Xu}, \citenamefont {Li}, \citenamefont {Xu},\ and\ \citenamefont {Wang}}]{Bin_2021}%
  \BibitemOpen
  \bibfield  {author} {\bibinfo {author} {\bibfnamefont {Z.-H.}\ \bibnamefont {Wang}}, \bibinfo {author} {\bibfnamefont {F.}~\bibnamefont {Xu}}, \bibinfo {author} {\bibfnamefont {L.}~\bibnamefont {Li}}, \bibinfo {author} {\bibfnamefont {D.-H.}\ \bibnamefont {Xu}}, \ and\ \bibinfo {author} {\bibfnamefont {B.}~\bibnamefont {Wang}},\ }\href {\doibase 10.1103/PhysRevB.104.174501} {\bibfield  {journal} {\bibinfo  {journal} {Phys. Rev. B}\ }\textbf {\bibinfo {volume} {104}},\ \bibinfo {pages} {174501} (\bibinfo {year} {2021}{\natexlab{c}})}\BibitemShut {NoStop}%
\bibitem [{\citenamefont {Longhi}(2021{\natexlab{b}})}]{longhi_2021}%
  \BibitemOpen
  \bibfield  {author} {\bibinfo {author} {\bibfnamefont {S.}~\bibnamefont {Longhi}},\ }\href {\doibase 10.1103/PhysRevB.103.224206} {\bibfield  {journal} {\bibinfo  {journal} {Phys. Rev. B}\ }\textbf {\bibinfo {volume} {103}},\ \bibinfo {pages} {224206} (\bibinfo {year} {2021}{\natexlab{b}})}\BibitemShut {NoStop}%
\bibitem [{\citenamefont {Zhou}(2021)}]{Zhou_2021}%
  \BibitemOpen
  \bibfield  {author} {\bibinfo {author} {\bibfnamefont {L.}~\bibnamefont {Zhou}},\ }\href {\doibase 10.1103/PhysRevResearch.3.033184} {\bibfield  {journal} {\bibinfo  {journal} {Phys. Rev. Res.}\ }\textbf {\bibinfo {volume} {3}},\ \bibinfo {pages} {033184} (\bibinfo {year} {2021})}\BibitemShut {NoStop}%
\bibitem [{\citenamefont {Zhou}\ and\ \citenamefont {Gu}(2022)}]{Zhou_2022}%
  \BibitemOpen
  \bibfield  {author} {\bibinfo {author} {\bibfnamefont {L.}~\bibnamefont {Zhou}}\ and\ \bibinfo {author} {\bibfnamefont {Y.}~\bibnamefont {Gu}},\ }\href {\doibase 10.1088/1361-648X/ac4530} {\bibfield  {journal} {\bibinfo  {journal} {Journal of Physics: Condensed Matter}\ }\textbf {\bibinfo {volume} {34}},\ \bibinfo {pages} {115402} (\bibinfo {year} {2022})}\BibitemShut {NoStop}%
\bibitem [{\citenamefont {Zeng}\ and\ \citenamefont {Xu}(2020)}]{Yong_2020}%
  \BibitemOpen
  \bibfield  {author} {\bibinfo {author} {\bibfnamefont {Q.-B.}\ \bibnamefont {Zeng}}\ and\ \bibinfo {author} {\bibfnamefont {Y.}~\bibnamefont {Xu}},\ }\href {\doibase 10.1103/PhysRevResearch.2.033052} {\bibfield  {journal} {\bibinfo  {journal} {Phys. Rev. Res.}\ }\textbf {\bibinfo {volume} {2}},\ \bibinfo {pages} {033052} (\bibinfo {year} {2020})}\BibitemShut {NoStop}%
\bibitem [{\citenamefont {Zhai}\ \emph {et~al.}(2022)\citenamefont {Zhai}, \citenamefont {Huang},\ and\ \citenamefont {Yin}}]{Zhai_2022}%
  \BibitemOpen
  \bibfield  {author} {\bibinfo {author} {\bibfnamefont {L.-J.}\ \bibnamefont {Zhai}}, \bibinfo {author} {\bibfnamefont {G.-Y.}\ \bibnamefont {Huang}}, \ and\ \bibinfo {author} {\bibfnamefont {S.}~\bibnamefont {Yin}},\ }\href {\doibase 10.1103/PhysRevB.106.014204} {\bibfield  {journal} {\bibinfo  {journal} {Phys. Rev. B}\ }\textbf {\bibinfo {volume} {106}},\ \bibinfo {pages} {014204} (\bibinfo {year} {2022})}\BibitemShut {NoStop}%
\bibitem [{\citenamefont {Gong}\ \emph {et~al.}(2018)\citenamefont {Gong}, \citenamefont {Ashida}, \citenamefont {Kawabata}, \citenamefont {Takasan}, \citenamefont {Higashikawa},\ and\ \citenamefont {Ueda}}]{Gong_2018}%
  \BibitemOpen
  \bibfield  {author} {\bibinfo {author} {\bibfnamefont {Z.}~\bibnamefont {Gong}}, \bibinfo {author} {\bibfnamefont {Y.}~\bibnamefont {Ashida}}, \bibinfo {author} {\bibfnamefont {K.}~\bibnamefont {Kawabata}}, \bibinfo {author} {\bibfnamefont {K.}~\bibnamefont {Takasan}}, \bibinfo {author} {\bibfnamefont {S.}~\bibnamefont {Higashikawa}}, \ and\ \bibinfo {author} {\bibfnamefont {M.}~\bibnamefont {Ueda}},\ }\href {\doibase 10.1103/PhysRevX.8.031079} {\bibfield  {journal} {\bibinfo  {journal} {Phys. Rev. X}\ }\textbf {\bibinfo {volume} {8}},\ \bibinfo {pages} {031079} (\bibinfo {year} {2018})}\BibitemShut {NoStop}%
\end{thebibliography}%
% \bibligraphystyle{ref}
\end{document}